# IS THERE A THERE THERE?
# TOWARD GREATER CERTAINTY FOR INTERNET JURISDICTION

Professor Michael Geist[*]

University of Ottawa, Faculty of Law

[*] The author would like to thank the Uniform Law Conference of Canada and Industry Canada for their financial support in sponsoring this paper; Teresa David and William Karam for their research assistance; as well as Vaso Maric, Rene Geist, and the participants at the Consumer Measures Committee/Uniform Law Conference of Canada April 2001 Workshop on Consumer Protection and Jurisdiction in Electronic Commerce for their comments on earlier versions of this paper. Any errors or omissions remain the sole responsibility of the author.

> "The Internet has no territorial boundaries. To paraphrase Gertrude Stein, as far as the Internet is concerned, not only is there perhaps "no there there," the "there" is everywhere where there is Internet access."[1]
> - Judge Nancy Gertner, *Digital Equipment Corp. v. Altavista Technology, Inc.*, 1997

> "We order the company YAHOO! Inc. to take all measures to dissuade and make impossible any access via Yahoo.com to the auction service for Nazi objects and to any other site or service that may be construed as constituting an apology for Nazism or contesting the reality of Nazi crimes..."[2]
> - Judge Jean-Jacques Gomez, *UEJF et LICRA v. Yahoo! Inc. et Yahoo France*, May 2000

Part I – Introduction

As business gravitated to the Internet in the late 1990s, concern over the legal risks of operating online quickly moved to the fore as the legal issues inherent in selling products, providing customer service, or simply maintaining an information-oriented Web site began to emerge.[3] Certain legal risks, such as selling defective products or inaccurate information disclosure were already well-known to business, since these risks are encountered and addressed daily in the offline world.[4]

The unique challenge presented by the Internet is that compliance with local laws is rarely sufficient to assure a business that it has limited its exposure to legal risk. Since Web sites are accessible worldwide, the prospect that a Web site owner might be hauled into a courtroom in a far-off jurisdiction is much more than a mere academic exercise – in an Internet environment that provides instant global access, it is a very real possibility.[5] For businesses seeking to embrace the promise of a global market at the click of a mouse, the prospect of additional compliance costs and possible litigation must be factored into the analysis.

---

[1] *Digital Equipment Corp. v. Altavista Technology, Inc.*, 960 F. Supp. 456 (D. Mass. 1997).
[2] *UEJF et LICRA v. Yahoo! Inc. et Yahoo France*, Tribunal De Grande Instance De Paris, N° RG: 00/05308, May 22, 2000.[hereinafter *LICRA v. Yahoo!*]
[3] *See*, *e.g.,* L. Trager, *Unhappy Holidays At Toys "R" Us,* ZDNet Interactive Week (12 January 2000), online: ZDNet <http://www.zdnet.com/filters/printerfriendly/0,6061,2421416-35,00.html> (date accessed: 30 March 2001).
[4] "Products Liability Law: An Overview" Cornell Law School, online: Cornell Law School <http://www.law.cornell.edu/topics/products_liability.html> (date accessed: 30 march 2001).
[5] *LICRA v. Yahoo!*, *supra*, and *Braintech, Inc. v. Kostiuk*, 1999 ACWSJ LEXIS 1924 (BCCA 1999) [hereinafter *Braintech*].

The risks are not limited to businesses, however. Consumers anxious to purchase online must also balance the promise of unlimited choice, greater access to information, and a more competitive, global marketplace with the prospect that they will not benefit from the security normally afforded by local consumer protection laws. Although such laws exist online just as they do offline, their effectiveness is severely undermined if consumers do not have recourse to their local court system or if enforcing a judgment requires a further proceeding in another jurisdiction.[6]

Moreover, concerns over the legal risks created by the Internet extend beyond commercial activities. Public interest information-based sites on controversial topics may face the prospect of prosecution in far-away jurisdictions despite the fact that the site may be perfectly legal within its home jurisdiction.[7] Anonymous posters to Internet chat sites, meanwhile, face the possibility that the target of their comment will launch a legal action designed chiefly to uncover their anonymous guise.[8]

---

[6] The U.S. Federal Trade Commission has noted:
> "Shifting to a pure country-of-origin approach to address challenges inherent in the current system risks undermining consumer protection, and ultimately consumer confidence in e-commerce. The same would be true under a "prescribed-by-seller" approach to the extent it would allow contractual choice-of-law and choice-of-forum provisions dictated by the seller to override the core protections afforded to consumers in their home country or their right to sue in a local court."

*Consumer Protection in the Global Electronic Marketplace: Looking Ahead (Staff Report),* US Federal Trade Commission - Bureau of Consumer Protection (September 2000) at 7, online: US Federal Trade Commission, <http://www.ftc.gov/bcp/icpw/lookingahead/electronicmkpl.pdf.> (date accessed: 30 March 2001).

[7] *LICRA v. Yahoo!, supra,* also see *Yahoo! ordered to bar French from Nazi sites*, Reuters (20 November 2000), online: ZDNet UK <http://www.zdnet.co.uk/news/2000/46/ns-19192.html> (date accessed: 30 March 2001).

[8] In Canada, *see, Irwin Toy Ltd. v. Doe* [2000] O.J. No. 3318. U.S. Cases include *J. Erik Hvide v. "John Does 1-8," et al.*, No. 99-22831-CA01, Fla. Cir., Miami-Dade Co.; *John Doe, also known as Aquacool_2000 v. Yahoo! Inc.*, Civ. Action No. 00-20677 (complaint filed in Super. Court for the State of California for the County of Los Angeles on May 11, 2000, subsequently transferred to San Jose). For a general review of the issue, *see*, C. S. Kaplan, J*udge Says Online Critic Has No Right To Hide*, New York Times Cyber Law Journal (9 June 2000), online: NY Times <http://www.nytimes.com/library/tech/00/06/cyber/cyberlaw/09law.html> (date accessed: 30 March 2001).

The challenge of adequately accounting for the legal risk arising from Internet jurisdiction has been exacerbated in recent years by the adoption of the *Zippo* legal framework,[9] commonly referred to as the passive versus active test. The test provides parties with only limited guidance and often results in detrimental judicial decisions from a policy perspective. As courts start to break free from the passive versus active test, they begin to shift toward an equally problematic effects-based approach that potentially grants jurisdiction to every court everywhere.[10]

Consider, for example, two Internet cases from 2000 with cross-border implications and challenging fact scenarios – the Yahoo.com France case[11] and the iCraveTV case.[12]

a.      The Yahoo.com France case

Few Internet law cases have attracted as much attention as the Yahoo! France case, in which a French judge ordered the world's most popular Web site to implement technical or access control measures blocking auctions featuring Nazi memorabilia hosted on the Yahoo.com site from French residents.[13] Yahoo! reacted with alarm, maintaining that the French court could not properly assert jurisdiction over the matter. It noted that the company maintains dozens of country-specific Web sites, including a

---

[9] *Zippo Manufacturing Co. v. Zippo Dot Com, Inc.*, 952 F. Supp. 1119 (W.D. Pa. 1997) [hereinafter *Zippo*].
[10] *See*, Part III, section B, *infra*.
[11] *LICRA v. Yahoo!, supra.*
[12] *Twentieth Century Fox Film Corp., et al. v. iCraveTV, et al.*, 2000 U.S. Dist. LEXIS 1013 (W.D. Pa. Jan. 28, 2000). [hereinafter *iCraveTV*]
[13] *See*, Jim Hu and Evan Hansen, *Yahoo Auction Case May Reveal Borders Of Cyberspace*, CNET News.com (11 August 2000), online: C-Net News <http://news.cnet.com/news/0-1005-200-2495751.html> (date accessed: 31 March 2001) ("A warning to Internet companies doing business abroad: Local governments may have the power to impose restrictions even if your servers are in the United States."); Kristi Essick, *Yahoo Told to Block Nazi Goods From French,* The Standard (20 November 2000), online, The Standard <http://www.thestandard.com/article/article_print/0,1153,20320,00.html> (date accessed: 31 March 2001) ("A French judge upholds his previous decision, ordering the company to install a filtering system for its auction site. The case raises questions about the jurisdiction of national courts over international Net companies.");

Yahoo.fr site customized for France that is free of Nazi-related content.[14] These country-specific sites target the local population in their local language and endeavour to comply with all local laws and regulations.

The company went on to argue that its flagship site, Yahoo.com, was targeted primarily toward a U.S. audience. Since Nazi memorabilia is protected under U.S. free speech laws, the auctions were entirely lawful there. Moreover, the Yahoo.com site featured a terms of use agreement which stipulated that the site was governed by U.S. law. Since the Yahoo.com site was not intended for a French audience and users implicitly agreed that U.S. law would be binding, the company felt confident that a French judge could not credibly assert jurisdiction over the site.[15]

Judge Jean-Jacques Gomez of the County Court of Paris disagreed, ruling that the he was entitled to assert jurisdiction over the dispute since the content found on the Yahoo.com site was available to French residents and was unlawful under French law.[16] Before issuing his final order, the judge commissioned an international panel to determine whether the technological means were available to allow Yahoo! to comply with an order to keep the prohibited content away from French residents. The panel reported that though such technologies were imperfect, they could accurately identify French Internet users at least seventy percent of the time.[17]

Based on that analysis, Judge Gomez ordered Yahoo! to ensure that French residents could not access content on the site that violated French law. Failure to comply with the order would result in fines of $13,000 per day.[18] Soon after, Yahoo! removed

---

[14] Brian Love, *Can Neo-Nazis Yahoo! in France?,* Reuters (10 August 2000), online: ZDNet News <http://www.zdnet.com/zdnn/stories/news/0,4586,2614196,00.html> (date accessed: 31 March 2001) ("French law prohibits the sale or exhibit of objects with racist overtones and none are directly available or visible on the Yahoo.fr site.")
[15] *Id*.
[16] *LICRA v. Yahoo!, supra.*
[17] *LICRA v. Yahoo! Inc.,* County Court of Paris, N° RG : 00/05308, May 2000 (Interim Court Order), online: Internet Societal Task Force <http://www.istf.org/archive/yahoo_france.html> (date accessed: 31 March 2001).
[18] *LICRA v. Yahoo!, supra.*

the controversial content from its site,[19] but the company proceeded to contest the validity of the French court's order in a California court.[20] That case is still pending.

b.     The iCraveTV case

In late 1999, iCraveTV, a small Canadian Internet startup company, attracted the legal wrath of broadcasters, sports leagues, and movie studios in both Canada and the U.S. when it began providing Internet users with the opportunity to watch television in real-time directly on their personal computers.[21] The lawsuits proved effective, as on February 28, 2000, approximately one month after it put a temporary stop to its webcasting activities under judicial order from a federal court in Pittsburgh,[22] iCraveTV announced that it had reached a settlement with the broadcasters, sports leagues, and movie studios on both sides of the border, agreeing to permanently stop its unauthorized webcasting activities.[23]

---

[19]L. Guernsey, *Yahoo to Try Harder to Rid Postings of Hateful Material*, New York Times (3 January 2001), online: NY Times
<http://www.nytimes.com/2001/01/03/technology/03YAHO.html?printpage=yes> (date accessed: 31 March 2001) ("Yesterday, Yahoo officials said the monitoring policy was not a response to the French ruling. Rather, they said, the company was responding to users who had requested a more active policy and to groups like the Wiesenthal Center and the Anti-Defamation League, which have been in talks with Yahoo throughout the year.").
[20] *Yahoo!, Inc. v. LICRA*, No. C 00-21275 (PVT) (N.D. Cal. Dec. 21, 2000).
[21] "A tiny Canadian Internet startup is being hit with the wrath of Hollywood and the big broadcasting networks in the U.S. The company is called iCraveTV. It's been in business less than two months, and it's just been hit with a huge lawsuit, backed by most of the American entertainment industry." Michael Colton, *U.S. Broadcasters Take iCraveTV To Court*, CBC Radio - World Report (21 January 2000), online: CBC Radio
<http://www.infoculture.cbc.ca/archives/newmedia/newmedia_01202000_icraveTVaction.phtml> (date accessed: 31 March 2001).
[22] *ICraveTV*, *supra*.
[23] Bloomberg News, *Broadcasters Pull Plug on iCraveTV*, *CNET News.com* (28 February 2000), online: CNET News.com <http://news.cnet.com/category/0-1004-200-1559907.html> (date accessed: 9 May 2000).
  For the settlement agreement, see online: Canadian Association of Broadcasters <http://www.cab-acr.ca/english/joint/submissions/settlement.htm> (date accessed: 10 May 2000).
  Interestingly, the settlement provides that if a court in Canada makes a final determination that Internet Webcasting without permission is not a violation of Canadian copyright law, iCrave can move to vary the terms of the settlement. Moreover, in addition to stopping the webcasting, iCrave agreed to stop its application for an Internet royalty before the copyright board.

One of the most interesting aspects of the case was the ease with which a U.S. court asserted jurisdiction over a Canadian company webcasting in Canada. iCraveTV had sought to limit its distribution to Canadians and thus avoid U.S. jurisdiction. Since iCraveTV recognized that its activities were legal in Canada but potentially illegal elsewhere, it conditioned access on passing through three stages of verifications and clickwrap agreements to ensure that only persons located in Canada could lawfully access the service. The first step required the potential user to enter their local area code. If the area code was not a Canadian area code, the user was denied access to the service. This approach was viewed, with some justification, as rather gimmicky since iCraveTV's own Toronto area code was posted on the site.[24]

The second step required the user to enter into a clickwrap agreement in which the user confirmed that they were located in Canada. The user was confronted with two icons -- an "In Canada" icon and a "Not in Canada" icon. If the user clicked on the "In Canada" icon, they were then presented with the third step, another clickwrap agreement. This agreement contained a complete terms of use agreement including another confirmation that the user was located in Canada. The user was required to scroll to the bottom of the agreement and click on the "I Agree" icon.

Since U.S. based users were required to pass through three stages to access the site, including fraudulently entering into two clickwrap agreements, it is arguable that under the *Zippo* test described below, while the iCraveTV site was "active" in Canada, it was actually passive for U.S. purposes and therefore should have fallen outside U.S. jurisdiction.

In light of the varied standards being applied by courts to establish jurisdictional rights in the online environment, this paper examines the effectiveness of the current approaches and recommends possible reforms. I argue that the passive versus active test established in *Zippo* has with time become increasingly outdated and irrelevant. It has been surpassed in practice by an effects-based analysis that poses even greater danger to legal certainty and the prospect for "over-regulation" of Internet-based activities. I argue instead for the adoption of a three-factor targeting test that includes analysis of contract, technology, and knowledge, as the standard for assessing Internet jurisdiction claims.

---

[24] See online: ICRAVETV.com <http://www.icravetv.com/adinfo/adinfo_frameset.html> (date accessed: 9 May 2000).

The paper is divided into five parts including the preceding introduction. Part two analyzes the complications created by Internet jurisdiction, highlighting four policy considerations that must be balanced in order to develop a test that both garners approval from a diverse group of stakeholders and remains relevant as technologies change - foreseeability, bias towards effects-based analysis, jurisdictional quid pro quo, and technological neutrality.

Part three reviews recent Internet jurisdiction jurisprudence in both the U.S. and Canada beginning with the development and subsequent approval for the *Zippo* passive versus active test. It then identifies the subtle changes that have been occurring since late 1999, as courts begin to find the test too constraining and shift their analysis toward an effects-based paradigm.

Part four presents the case for a targeting-based test for Internet jurisdiction, demonstrating first the growing support for targeting in both the case law and at the international policy level. It then advocates for the adoption of a three-factor approach to targeting that includes assessments of any contractual provisions related to jurisdiction, the technological measures employed to identify the targeted jurisdiction, and the actual or implied knowledge of the Web site operator with respect to targeted jurisdictions.

Part five concludes by applying the targeting test to the Yahoo! France and iCraveTV cases. Although the analysis would not change the outcome in these cases, it demonstrates how the parties would benefit from greater legal certainty in applying a targeting-based analysis.

Part II – Jurisdiction on the Internet

Professor Yochai Benkler of the NYU Law School argues that communications systems are divided among three interconnected layers.[25] There is a physical layer that includes the physical wires and connections needed to connect phones, computers, routers, and other communicating technology. Above that there is a logical layer that determines who is able to access what on the network. And finally, above that there is a content layer where the content being communicated resides.

Internet jurisdiction can also be conceptualized in three layers. There is an application layer that determines whether courts are entitled to apply their laws to a particular dispute. Above that there is a substantive layer where courts apply their substantive laws to the dispute.[26] And above that layer is the enforcement layer, where court orders must be enforced in an online environment that often resists the imposition of foreign judgments due to large distances and small monetary disputes.[27]

Internet jurisdiction discussions often fail to adequately distinguish between these three layers. For example, criticism leveled at the French court's decision in the Yahoo!

---

[25] *A Free Information Ecology In The Digital Environment (New York University, Conference Session 12),* The Information Law Institute at New York University School of Law, at 29, online: New York University School of Law <http://www.law.nyu.edu/ili/conferences/freeinfo2000/webcast/transcripts/105124DemDiscourse.pdf> (date: accessed: 30 March 2001).

[26] The substantive layer tends to be the most contentious since it frequently pits divergent perspectives on fundamental legal freedoms such as freedom of speech. *See*, *e.g.,* Stephan Wilske and Teresa Schiller, *International Jurisdiction In Cyberspace: Which States May Regulate The Internet?*, 50 Fed. Comm. L.J. 117, 122-3 (1997) ("When CompuServe, Inc. blocked access by its subscribers in the United States and around the world to two hundred discussion groups after a federal prosecutor in Germany had indicated that they might violate German pornography laws, users realized that 'cyberspace doesn't belong to a single country,' but to a whole range of countries with diverse legal concepts.")

[27] Despite the focus on the application layer, better known as adjudicatory jurisdiction, some commentators have opined that the enforcement layer actually presents the greatest challenge in the online environment. *See*, *e.g*., Henry H. Perritt, Jr., *Will the Judgment-Proof Own Cyberspace?,* 32 Int'l Law. 1121, 1123 (1998) [hereinafter *Perritt*] ("The real problem is turning a judgment supported by jurisdiction into meaningful economic relief. The problem is not the adaptability of International Shoe-obtaining jurisdiction in a theoretical sense. The problem is obtaining meaningful relief.").

case has focused on the court's willingness to assert jurisdiction over a U.S. based site, the inappropriateness of French free speech law, and the challenge of forcing Yahoo! to comply with the order.[28] Although each of these criticisms is treated as a single critique of the case, in fact, each involves a separate jurisdictional layer and merits a different response.

This paper focuses on Internet jurisdiction's application layer. Arguments over the substantive layer are much more difficult defend – different countries have different norms and principles and it is unrealistic to expect the Internet to spur harmonization of all substantive issues. Similarly, arguments over the enforcement layer tend to involve business risk analysis, rather than legal risk analysis, since the ability to enforce a local decision will often depend upon whether the affected party has local assets subject to seizure or is sufficiently large that it cannot afford to ignore an outstanding court order, no matter where it is located.[29]

Sorting through conflicting laws and competing claims often presents lawmakers and courts with several difficult policy choices - choices that tend to blur the distinction between the three layers. For example, although it might be intuitively argued that local laws should protect consumers online in the same manner as they protect consumers offline, the application of these principles on the Internet is particularly complex. This complexity raises concern over the application of local law, the desire to protect local citizenry from harmful cyber-effects, the furtherance of policy goals that seek to encourage e-commerce and Internet use, as well as the difficulty in defining policies that

---

[28] The case was characterized in the following manner by the Center for Democracy and Technology:
> "In a setback for free expression on the Internet, a French court has ruled that U.S.-based Yahoo, Inc. is to be held liable under French law for allowing French citizens to access auction sites for World War II Nazi memorabilia. ... The ruling appears to impose blocking requirements that many view as impractical to implement on a wide scale and highly imperfect at identifying Internet users by country. It also sets a dangerous precedent for countries seeking to impose restrictions on speech outside their borders."

*French Court Holds Yahoo Accountable For U.S. Auction Content*, Center For Democracy and Technology, online: CDT <http://www.cdt.org/publications/pp_6.20.shtml> (date accessed: 31 March 2001).
[29] *LICRA v. Yahoo!*, supra.

can be applied in a technology neutral fashion. Since the issues to be considered capture elements of all three layers one must ask – should local courts assert jurisdiction over every online consumer purchase (application layer)? Should policy explicitly seek to encourage e-commerce by creating e-commerce specific consumer protection legislation (substantive layer)? Will a consumer actually benefit from a local judgment if the award must still be enforced elsewhere (enforcement layer)?

At the heart of the application layer lies a deceptively simple question – when is it appropriate to assert jurisdiction over Internet-based activities? Since the question of jurisdiction is not new (most countries have a rich body of law addressing conflict of laws, choice of forum, and enforcement of judgments),[30] most courts and policy makers quite properly revert to first principles to develop appropriate guidelines.[31]

In many jurisdictions, the litmus test for determining whether assertion of jurisdiction is appropriate involves analyzing whether jurisdiction is reasonable under the circumstances. While admittedly a subjective concept, courts in the U.S and Canada have regularly relied on a reasonableness standard as their guide. In the U.S., the reasonableness standard is couched in terms of "minimum contacts",[32] while in Canada the language of choice is "real and substantial connection."[33] Although these terms necessitate somewhat different analyses, the core principle remains the same – that the appropriateness of asserting jurisdiction depends upon whether the parties themselves think it reasonable to do so.

Unfortunately, aside from reassuring parties that jurisdiction will not be asserted in an indiscriminate manner, substituting the word "reasonable" for "appropriate" does little to provide much additional legal certainty. Accordingly, it has fallen to the case law to provide guidance on how the term "reasonable" should be interpreted. Case law analysis suggests that within the context of jurisdiction law, a foreseeability metric lies at the heart of the reasonableness which dictates that a party should only be hauled into a foreign court where it was foreseeable that such an eventuality might occur.

---

[30] Eugene F. Scoles and Peter Hay. *Conflict of Laws (Hornbrook Series* (Minnesota: West Publishing Company, December 1998).
[31] Ogilvy Renault Internet Group, *Jurisdiction and the Internet: are Traditional Rules Enough?*, (Canada: Uniform Law Conference of Canada, July 1998), online: Uniform Law Conference of Canada <http://www.law.ualberta.ca/alri/ulc/current/ejurisd.htm> (date accessed: 30 March 2001).
[32] *International Shoe Co. v. Washington*, 326 U.S. 310, 66 S.Ct. 154, 90 L.Ed. 95 (1945).
[33] *Morguard Investments Ltd. v. De Savoye* [1990] 3 S.C.R. 1077.

Although a foreseeability test may not always provide absolute legal certainty, it does provide an intuitive sense of when a court will assert jurisdiction over a dispute. For example, if a contract dispute arises between two parties in different countries, it would generally be considered foreseeable that, absent a forum selection clause (in which case the parties have settled on the governing jurisdiction in advance of the dispute), courts in either country might be willing to assert jurisdiction. In other instances, such as a defamation tort action, a court would likely conduct an effects-based analysis based on foreseeability, concluding that the alleged defamer would have foreseen that the defamatory statements would have an impact within the defamed party's jurisdiction and thus they might face the prospect of litigation there.

While the foreseeability/reasonableness standard may have functioned effectively in the offline world, there are several reasons why the Internet complicates the issue. First, with worldwide Internet availability,[34] foreseeability is much more difficult to gauge. Scholars have commented that the "borderless Internet" significantly impedes the application of physical laws, leading some to advocate for a separate cyberspace jurisdiction.[35] Since the test is rooted in the principle of providing greater clarity, the Internet clouds matters by providing an "all or nothing" environment in which either every jurisdiction is foreseeable or none is.

Second, courts and policy makers are likely to bias toward asserting jurisdiction where harm has been experienced locally.[36] This can best be understood by considering a simple business-to-consumer e-commerce transaction. Consider a consumer located in Ottawa who downloads an electronic book from Amazon.com, a leading e-commerce business located in Seattle, Washington. If the consumer is dissatisfied with the transaction – the downloaded e-book causes his computer system to crash and lose valuable data – and the parties are unable to negotiate a settlement, the consumer may wish to sue for the price of the book and resulting damages in a local Ontario court. Amazon is likely to contest the action on jurisdictional grounds, arguing that the sales contract between the parties stipulates that any disputes should be brought in a court in Washington state.

---

[34] See Part IV, section 2.
[35] David R. Johnson and David G. Post, *Law And Borders--The Rise of Law in Cyberspace*, Stan. L. Rev. 1367 (1996) [hereinafter *Law and Borders*].
[36] According to one report on European activity:

Should the Ontario court dismiss the action by upholding the enforceability of the forum selection clause? Will doing so effectively eliminate the consumer's access to Ontario consumer protection legislation? Courts throughout North America appear divided on the issue. In a 1999 Ontario case, a court dismissed a class action lawsuit brought against Microsoft on the basis that a clickwrap agreement between the parties provided for the state of Washington law to govern any dispute.[37] A recent California case ruled in the opposite manner, however, holding that a dispute between AOL and one of its customers could be heard in a California court despite the existence of a forum selection clause that provided that all disputes be brought in a Virginia court.[38]

---

"The law, dubbed the Brussels I regulation, will come into effect next March. It states that where there is a dispute between a consumer in one EU country and an online retailer in another, the consumer will be able to sue in a court in his own country. The justice ministers and the European Commission, which drafted the regulation, argue that this focus on the consumer is essential to help get electronic commerce off the ground in Europe. "A lack of consumer confidence is the main thing holding up the development of e-commerce here," said Leonello Gabrici, spokesman on judicial matters for the Commission. He said that by handing jurisdiction of such cross-border disputes to the courts in the consumers' country, the regulation will be encouraging consumers to purchase online."
Paul Meller, *European Justices Pass Stiff E-Commerce Law*, IDG (30 November 2000), online: IDG <http://www.idg.net/ic_300048_1794_9-10000.html> (date accessed: 31 march 2001).
[37] *Rudder v. Microsoft Corporation* (1999), 2 C.P.R. (4th) 474 (Ont. S.C.J.).
[38] *Mendoza v. AOL*, Superior Ct. of Cal., County of Alameda, Dept. No. 22. (unreported, on file with the author) [hereinafter *Mendoza*].

The scenario becomes even more complicated when the case involves free speech rather than commercial concerns. For example, the recent French Nazi memorabilia case involving Yahoo! illustrates how a local court may be willing to assert jurisdiction even in the absence of evidence that the harm was directed at the jurisdiction, reasoning that the perceived local harm is too great to ignore.[39] While such an approach raises few concerns when it involves activity such as securities fraud where global rules are relatively uniform,[40] the application of an effects-based standard to issues such as free speech is likely to prove highly contentious.

Although courts and policy makers may have a bias towards protecting local citizenry from commercial or content harm, the issue is even further complicated by the fact that all countries face the same concern. Accordingly, while a country may wish to protect its own consumers by asserting jurisdiction over out-of-country entities, it would prefer that other countries not exert the same authority over its citizens and companies.[41]

Moreover, the laws being applied locally will vary since different countries will promote different policy priorities. Some countries may view consumer protection as more important than promotion of e-commerce growth and thus adopt a policy of aggressively asserting jurisdiction to protect local consumers. Others may favour the promotion of privacy protection and will thus seek to assert its jurisdiction over a privacy framework. As Lawrence Lessig argues in his seminal book, Code and Other Laws of Cyberspace, these competing policy priorities encourage countries to engage in a quid pro quo approach to jurisdictional cooperation.[42] In discussing the State of Minnesota's desire to enforce state anti-gambling laws, Lessig notes:

> Why would any other jurisdiction want to carry out Minnesota's regulation?
>
> The answer is that they would not if this were the only regulation at stake. Minnesota wants to protect its citizens from gambling, but New York may want to protect its citizens against the misuse of private data. The European Union may share New York's objective; Utah may share Minnesota's.
>
> Each state has its own stake in controlling certain behaviors, and these behaviors are different. But the key is this: the same architecture that enables Minnesota to achieve its regulatory end can also help other states achieve their regulatory ends. And this can initiate a kind of quid pro quo between jurisdictions.[43]

---

[39] *LICRA v. Yahoo!, supra.*
[40] *Securities Activity on the Internet*, International Organisation of Securities Commissions (IOSCO) - Technical Committee (September 1998), online: <http://www.iosco.org/download/pdf/1998-internet_security.pdf> (date accessed: 1 April 2001) [hereinafter *IOSCO*].
[41] Dean Henry Perritt notes that "Extending the bases of jurisdiction is a two-edged sword. United States citizens may be able to assert U.S. law in U.S. courts with respect to harmful conduct occurring offshore, but they also may be subject to prosecution or litigation in foreign tribunals."
*Perritt*, *supra* at 1131.

As if the policy choices weren't already sufficiently complicated, an additional consideration must be factored into the analysis. As policy makers continue to grapple with the challenges of the Internet, it has become increasingly accepted that establishing effective and enduring guidelines or standards for the Internet requires the adoption of a "technology neutral" approach.[44] Technology neutral approaches have been a hallmark of many Internet law policy initiatives, including the development of e-commerce legislation in Canada[45] and the adoption of electronic evidence statutes.[46] Technology neutral in this context refers to statutory tests or guidelines that do not depend upon a specific development or state of technology, but rather are based on core principles that can be adapted to changing technologies. Since technological change is constant, standards created with specific technologies in mind are likely to become outdated as the technology changes. Applied to the Internet jurisdiction context, using indicia that reflect the current state of the Internet and Internet technologies is a risky proposition, since those indicia risk irrelevancy when the technology changes.

---

[42] Lawrence Lessig, *Code And Other Laws Of Cyberspace*, at 55 (Basic Books 1999) [hereinafter *Code*].
[43] *Ibid.*
[44] As the Australian Attorney General's office has noted within the context of UNCITRAL e-commerce negotiations:
  "A technology neutral approach is preferable as it has become clear that technology specific legislative schemes can inhibit market choice. Furthermore, legislative regimes that prefer one technology over another create impediments to electronic commerce and restrict innovation."
*UNCITRAL Developments*, Australian Attorney General's Department - Information Economy Section, online: Australian Attorney General's Department <http://law.gov.au/publications/ecommerce/> (date accessed: 30 March 2001).
[45] *Uniform Electronic Commerce Act* (Model Law), Uniform Law Conference of Canada (ULCC), online: ULCC <http://www.law.ualberta.ca/alri/ulc/current/euecafin.htm> (date accessed: 30 March 2001).
[46] *Uniform Electronic Evidence Act* (Model Law), Uniform Law Conference of Canada (ULCC), online: ULCC <http://www.law.ualberta.ca/alri/ulc/current/eeeact.htm> (date accessed: 30 March 2001).

In seeking to balance the four factors – foreseeability, bias towards effects-based analysis, jurisdictional quid pro quo, and technological neutrality – the development of a single standard for Internet jurisdiction analysis presents a difficult though not insurmountable challenge. Unfortunately, the current passive versus active test nudges the law squarely in the wrong direction by failing to provide parties with sufficient guidance on any of these four factors.

Part III – The Rise and Fall of the *Zippo* Test

A.     The Emergence of the *Zippo* Passive versus Active Test

Internet jurisdiction case law in North America traces back to 1996 and *Inset Systems, Inc. v. Instruction Set, Inc.*, a Connecticut District Court case.[47] In this instance, Inset Systems, a Connecticut company, brought a trademark infringement action against Instruction Set, a Massachusetts company, arising out of its use of the domain name, "Inset.com".[48] Instruction Set used the domain name to advertise its goods and services on the Internet, a practice to which Inset objected since it was the owner of the federal trademark "Inset". The legal question before the court was one of jurisdiction - did Instruction Set's activity, in this case the establishment of a Web site, properly bring it within the jurisdiction of Connecticut under that state's long-arm statute and by meeting the minimum contacts standard established by the U.S. Supreme Court in *World-Wide Volkswagen Corp. v. Woodson*?[49]

---

[47] *Inset Systems, Inc. v. Instruction Set, Inc.*, 937 F. Supp. 161 (D. Conn. 1996) [hereinafter *Inset Systems*].
[48] Internet domain names, which have become a ubiquitous part of commercial advertising, enable users to access Web sites simply by typing in a name such "www.inset.com" in their Web browser. The "www" portion of the address identifies that the site is part of the World Wide Web; the "Inset" portion is usually the name of a company or other identifying words; and "com" identifies the type of institution, in this case a company. Domain names, the subject of several other litigated cases, are administered in the United States by a government appointed agency, Network Solutions Inc. (NSI) and are distributed on a first come, first served basis.

The Court concluded that it could properly assert jurisdiction, basing its decision on Instruction Set's use of the Internet. Likening the Internet to a continuous advertisement, the Court reasoned that Instruction Set had purposefully directed its advertising activities toward Connecticut on a continuous basis by virtue of its Web site and, therefore, could reasonably have anticipated being hauled into court there.[50]

The Court's decision was problematic for several reasons. First, its conclusion that anyone who creates a Web site is purposefully directing their activity toward every jurisdiction stretches the meaning of "purposefully directing" activity to its outer limits. Second, the Court did not engage in any analysis of the Internet itself, but rather found it sufficient to analogize the Internet to a more traditional media form, in this case a continuous advertisement, and apply the existing law. However, the implications of its reasoning if legally correct - that in effect *every court anywhere* could assert jurisdiction on the basis that a Web site is directed toward that jurisdiction - could stifle future Internet growth since would-be Internet participants would be forced to weigh the advantages of the Internet with the potential of being subject to every legal jurisdiction in the world.

Third, the Court did not assess Instruction Set's actual activity on the Internet. The mere *use* of the Internet was sufficient for this court to establish jurisdiction. In fact, the Court acknowledged that Instruction Set did not maintain an office in Connecticut, nor did it have a sales force or employees in the state.[51] A more complete analysis of the underlying facts would have included an assessment of precisely what was happening on the Internet. Was Instruction Set selling products directly to people in Connecticut through its Web site? Was it providing a service directly through its Web site? Was it actively soliciting the participation of potential users by encouraging correspondence? What was the approximate number of Connecticut users who actually accessed the Web site? Asking these questions and others like them would have provided the Court with a much stronger basis for asserting that Instruction Set had purposefully directed its activity toward Connecticut and, moreover, would have developed a framework so that all Internet activity would not be viewed as equivalent.

---

[49] *World-Wide Volkswagen Corp. v. Woodson*, 444 U.S. 286 (1980).
[50] *Inset* Systems, *supra* at 165.
[51] *Id.*

With the *Inset* precedent established, many similar cases soon followed. In *Maritz, Inc. v. Cybergold, Inc.*, an August 1996 Missouri District Court case, the Court was again faced with the question of personal jurisdiction in the context of a trademark infringement action.[52] Citing with approval the *Inset Systems* decision, the Court struggled for an effective metaphor, noting that:

> "...the nature and quality of contacts provided by the maintenance of a web-site on the Internet are clearly of a different nature and quality than other means of contact with a forum such as the mass mailing of solicitations into a forum...or that of advertising an 800 number in a national publication."[53]

Unable to arrive at an effective analogy, the Court proceeded to conclude that it was a conscious decision to transmit advertising information to all Internet users and that such knowledge was sufficient for the assertion of personal jurisdiction.[54]

The Canadian experience on the matter of Internet jurisdiction closely mirrored that of the U.S. In keeping with the *Inset* line of cases, in *Alteen v. Informix Corp.,* a 1998 Newfoundland case, Judge Woolridge of the Newfoundland Supreme Court asserted jurisdiction based strictly on the provision of information via the Internet.[55]

The case involved allegations that Informix Corp., a U.S.-based maker of data storage media, issued false and misleading statements that led to an inflated stock price. When shareholders residing in Newfoundland launched a tort lawsuit, Informix responded by arguing that the local court could not properly assert jurisdiction since the company did not trade shares on a Canadian public exchange, issue public statements to the Canadian press, or maintain direct contact with the plaintiffs.

---

[52] *Maritz, Inc. v. Cybergold, Inc.*, 947 F. Supp. 1328 (E.D. Mo. 1996).
[53] *Id.* at 1332.
[54] *Id.*
[55] *Alteen v. Informix Corp.* [1998] N.J. No. 122 1997 No. C.B. 439 (Newf. S.C. - T.D.).

The court rejected the argument, siding with the plaintiffs who maintained that the availability of public statements on sources such as the Internet often led to Canadian media coverage. Since the shares were purchased in Newfoundland and corporate information was available within the province, the court ruled that it was entitled to assert jurisdiction over the tort action.

Although the action involved tort rather than trademark infringement, the Informix case bears a striking similarity to the early U.S. Internet cases, where the mere availability of information on the Internet was viewed as sufficient to assert jurisdiction. Had that analysis been adopted, all Canadian courts would theoretically be entitled to assert jurisdiction over parties posting information on the Internet.

While several additional U.S. cases did follow the *Inset* approach,[56] one New York District Court case stands out as an important exception.[57] "The Blue Note" was a small Columbia, Missouri club operated by Richard King. King promoted his club by establishing a Web site that included information about the club, a calendar of events, and ticketing information. New York City was also home to a club named "The Blue Note", this one operated by the Bensusan Restaurant Corporation, who owned a federal trademark in the name. King was familiar with the New York Blue Note as he included a disclaimer on his Web site that stated: "The Blue Note's Cyberspot should not be confused with one of the world's finest jazz club[s] [the] Blue Note, located in the heart

---

[56] *See, e.g.*, *Heroes, Inc. v. Heroes Foundation*, 958 F. Supp. 1 (D. D.C. 1996) (citing *Inset Systems* with approval in finding that a Web site sustained contact with the District of Columbia); and *Panavision International, L.P. v. Toeppen*, 938 F. Supp. 616 (C.D. Cal. 1996) (finding that use of a trademark infringing domain name in Illinois was an act expressly directed at California).
[57] *Bensusan Restaurant Corporation v. King*, 937 F. Supp. 295 (S.D.N.Y. 1996), *aff'd* 126 F. 3d. 25 (2nd Cir. 1997).

of New York's Greenwich Village. If you should find yourself in the big apple give them a visit".[58]

Within months of the establishment of King's Blue Note Web site, Bensusan brought a trademark infringement and dilution action in New York federal court. Once again, the Court faced the question of personal jurisdiction in the context of a trademark action arising out of activity on the Internet. Unlike the *Inset* line of cases, however, the Court here considered the specific uses of the Web site in question. It noted that King's Web site was passive rather than active in nature -- several affirmative steps by a New York resident would be necessary to bring any infringing product into the state. Specifically, tickets could not be ordered online, so that anyone wishing to make a purchase would have to telephone the box office in Missouri, only to find that the Missouri club did not mail tickets. The purchaser would have to travel to Missouri to obtain the tickets.[59] Given the level of passivity, the Court ruled that the Web site was not causing any infringing activity in New York.[60] In fact, the Court argued that "[T]he mere fact that a person can gain information on the allegedly infringing product is not the equivalent of a person advertising, promoting, selling or otherwise making an effort to target its product in New York."[61]

The *Bensusan* decision, which a Second Circuit court affirmed on appeal in September 1997, provided an important step toward the development of deeper legal analysis of Internet activity. Although the decision did not attempt to reconcile the *Inset* line of cases, it provided the groundwork upon which a new line of cases was advanced.[62]

---

[58] *Inset* Systems, *supra* at 297-8
[59] *Id.* at 299.
[60] *Id.*
[61] *Id.*
[62] For example, the February 1997 New York District Court decision of *Hearst Corporation v. Goldberger*, 1997 WL 97097 (S.D.N.Y. 1997) relied heavily upon the Bensusan analysis in refusing to assert personal jurisdiction in a trademark infringement matter involving the domain name "Esqwire.com". The *Goldberger* court carefully reviewed Internet case law to that point, noted its disagreement with decisions such as *Inset*, *Maritz*, *EDIAS*, and *Panavision*, and cautioned that:

However, by the end of 1996, the majority of Internet-related decisions evidenced little genuine understanding of activity on the Internet. Rather, most courts were unconcerned with the jurisdictional implications of their rulings and instead favored an analogy-based approach in which the Internet was categorized *en masse*.

In early 1997 a new approach emerged, led by the Pennsylvania District Court decision, *Zippo Manufacturing Co. v. Zippo Dot Com, Inc.*[63] It was with this decision that courts gradually began to appreciate that activity on the Internet was as varied as that in real space and that all-encompassing analogies could not be appropriately applied to this new medium.

Zippo Manufacturing was a Pennsylvania based manufacturer of the well-known "Zippo" brand of tobacco lighters. Zippo Dot Com was a California based Internet news service that used the domain name "Zippo.com" to provide access to Internet newsgroups. Dot Com offered three levels of subscriber service -- free, original, and super. Those subscribers desiring the original or super level of service were required to fill out an online application form and submit a credit card number through the Internet or by telephone. Dot Com's contacts with Pennsylvania occurred almost exclusively on the Internet since the company maintained no offices, employees, or agents in the state. Dot Com had some success in attracting Pennsylvania subscribers. At the time of the action, approximately 3,000 or two percent of its subscribers resided in that state. Once again, the issue before the court was one of personal jurisdiction arising out of a claim of trademark infringement and dilution.

---

> "Where, as here, defendant has not contracted to sell or actually sold any goods or services to New Yorkers, a finding of personal jurisdiction in New York based on an Internet web site would mean that there would be nationwide (indeed, worldwide) personal jurisdiction over anyone and everyone who establishes an Internet web site. Such nationwide jurisdiction is not consistent with traditional personal jurisdiction case law nor acceptable to the Court as a matter of policy."

[63] *Zippo, supra*.

Rather than using Internet analogies as the basis for its analysis, the Court focused on the prior, somewhat limited Internet case law.[64] The Court's examination of the case law, which clearly used the *Bensusan* decision for inspiration, yielded the following conclusion:

> "With this global revolution looming on the horizon, the development of the law concerning the permissible scope of personal jurisdiction based on Internet use is in its infant stages. The cases are scant. Nevertheless, our review of the available cases and materials reveals that the likelihood that personal jurisdiction can be constitutionally exercised is *directly proportionate to the nature and quality of commercial activity that an entity conducts over the Internet*. This sliding scale is consistent with well developed personal jurisdiction principles. At one end of the spectrum are situations where a defendant clearly does business over the Internet. If the defendant enters into contracts with residents of a foreign jurisdiction that involve the knowing and repeated transmission of computer files over the Internet, personal jurisdiction is proper. At the opposite end are situations where a defendant has simply posted information on an Internet Web site which is accessible to users in foreign jurisdictions. A passive Web site that does little more than make information available to those who are interested in it is not grounds for the exercise of personal jurisdiction. The middle ground is occupied by interactive Web sites where a user can exchange information with the host computer. In these cases, the exercise of jurisdiction is determined by examining the level of interactivity and commercial nature of the exchange of information that occurs on the Web site.[65]

Although the Court may have conveniently interpreted some earlier cases to obtain its desired result, its critical finding was that the jurisdictional analysis in Internet cases should be based on, as the Court states, the nature and quality of the commercial activity conducted on the Internet. There was a strong argument that prior to *Zippo*, the jurisdictional analysis was based upon the mere use of the Internet itself, a finding that might easily produce a somewhat inappropriate analogy and lead to the subsequent application of legal doctrine unsuited to the circumstances. In the aftermath of the *Zippo* decision, in which the Court used its analysis to find that jurisdiction was proper due to

---

[64] One case omitted from the discussion but relied upon by the *Zippo* court was *Compuserve, Inc. v. Patterson*, 89 F. 3d 1257 (6th Cir. 1996). Although the *Zippo* court refers to the decision as an Internet case, in fact, the activity in question did not involve the use of the Internet. Rather, Patterson used Compuserve's proprietary network to distribute certain shareware programs. Accordingly, Patterson's contacts with Ohio, Compuserve's headquarters and the location of the litigation, were confined to an off-line contractual agreement and the posting of shareware on a Compuserve server that was available to users of its proprietary network (not Internet users at large).

[65] 952 F. Supp. at 1124 (citations omitted)(emphasis added)

Dot Com's subscription sales to state residents, Internet legal analysis underwent a significant shift in perspective.

In the years since *Zippo*, the passive versus active approach has been cited with approval in numerous cases.[66] For example, in *Cybersell, Inc. v. Cybersell, Inc.*, a

---

[66] See also *Amberson Holdings LLC v. Westside Story Newspaper*, 110 F.Supp.2d 332 (U.S.D.C., D. New Jersey, 2000); *Hasbro, Inc. v. Clue Computing, Inc*., 66 F.Supp.2d 117 (D. Mass., 1999); *Search Force Inc. v. DataForce Intern.*, Inc., 112 F.Supp.2d 771 (S.D. Ind., 2000); *American Eyewear, Inc. v. Peeper's Sunglasses and Accessories*, Inc., 106 F.Supp.2d 895 (N.D. Tex., 2000); *Patriot Systems, Inc. v. C-Cubed Corp.*, 21 F.Supp.2d 1318 (D. Utah, 1998); *Roche v. Worldwide Media, Inc.*, 90 F.Supp.2d 714 (E.D. Va., 2000); *Coastal Video Communications, Corp. v. Staywell Corp.*, 59 F.Supp.2d 562, (E.D. Va., 1999); *Soma Medical Intern. v. Standard Chartered Bank*, 196 F.3d 1292 (10[th] Cir. (Utah), 1999); *Butler v. Beer Across America*, 83 F.Supp.2d 1261 (N.D. Ala., 2000); *Colt Studio, Inc. v. Badpuppy Enterprise*, 75 F.Supp.2d 1104 (C.D. Cal., 1999); *CIVIX-DDI LLC v. Microsoft Corp.*, 1999 WL 1020248 (D. Colo., 1999); *Blumenthal v. Drudge*, 992 F.Supp 44 (D.D.C., 1998); *J.B. Oxford Holdings, Inc. v. Net Trade, Inc.* 76 F.Supp.2d 1363 (S.D. Fla., 1999); *Berthold Types Ltd. v. European Mikrograf Corp.*, 102 F.Supp.2d 928 (N.D. Ill., 2000); *International Star Registry of Illinois v. Bowman-Haight Ventures, Inc.*, 1999 WL 300285 (N.D. Ill., 1999); *Vitullo v. Velocity Powerboat, Inc.*, 1998 WL 246152 (N.D. Ill., 1998); *F. McConnell and Sons, Inc. v. Target Data Systems, Inc.*, 84 F.Supp.2d 961 (N.D. Ind., 1999); *Resuscitation Technologies, Inc. v. Continental Health Care Corp.*, 1997 WL 148567 (S.D. Ind., 1997); *Alantech Distribution, Inc. v. Credit General Ins. Co.*, 30 F.Supp.2d 534 (D. Md., 1998); *McRae's, Inc. v. Hussain*, 105 F.Supp.2d 594 (S.D. Miss., 2000); *Lofton v. Turbine Design, Inc.*, 100 F.Supp.2d 404 (N.D. Miss., 2000); *Citigroup v. City Holding Co.*, 97 F.Supp.2d 549 (S.D.N.Y., 2000); *K.C.P.L., Inc. v. Nash*, 1998 WL 82367 (S.D.N.Y., 1998); *Tech Heads, Inc. v. Desktop Service Center, Inc.*, 105 F.Supp.2d 1142 (D.Or., 2000); *Standard Knitting, Ltd. v. Outside Design, Inc.*, 2000 WL 804434 (E.D. Pa., 2000); *Westcode, Inc. v. RBE Electronics, Inc.*, 2000 WL 124566 (E.D. Pa., 2000); *Harbuck v. Aramco, Inc.*, 1999 WL 999431 (E.D. Pa., 1999); *Renick v. Manfredy*, 52 F.Supp.2d 462 (E.D. Pa., 1999); *Barrett v. Catacombs Press*, 44 F.Supp.2d 717 (E.D. Pa., 1999); *Blackburn v. Walker Oriental Rug Galleries, Inc.*, 999 F.Supp 636 (E.D. Pa., 1998); *Brown v. Geha-Werke GmbH*, 69 F.Supp.2d 770 (D.S.C., 1999); *ESAB Group, Inc. v. Centricut, LLC*, 34 F.Supp.2d 323 (D.S.C., 1999); *Miecskowski v. Masco Corp.*, 997 F.Supp. 782 (E.D. Tex., 1998); *Agar Corp. Inc. v. Multi-Fluid Inc.*, 1997 WL 829340 (S.D. Tex., 1997); *Jewish Defence Organization, Inc. v. Superior Court*, 85 Cal. Rptr. 2d 611 (Cal.App. 2 Dist., 1999); *Nida Corp. v. Nida*, 2000 WL 1610635 (M.D. Fla., 2000); *Smith v. Hobby Loby Stores, Inc.,* 968 F.Supp. 1356 (W.D. Ark., 1997); *Online Partners.Com, Inc. v. Atanticet Media Corp.,* 2000 WL 101242 (N.D. Cal., 2000); *Quokka Sports, Inc., v. Cup Intern. Ltd.,* 99 F.Supp.2d 1105 (N.D. Cal., 1999); *Mallinkrodt Medical, Inc. v. Sonus Pharmaceuticals, Inc.*, 989 F.Supp 265 (D.D.C., 1998); *Ameritech Services, Inc. v. SCA Promotions, Inc.,* 2000 WL 283098 (N.D. Ill., 2000); *LFG, LLC v. Zapata Corp.,* 78 F. Supp.2d 731 (N.D. Ill., 1999); *Biometics, LLC v. New Womyn*, Inc., 112 F.Supp.2d 869 (E.D. Mo., 2000); *Decker v. Circus Circus Hotel*, 49 F.Supp.2d 748 (D.N.J., 1999); *Hurley v. Cancun Playa Oasis Intern. Hotels*, 1999 WL 718556 (E.D. Pa., 1999); *Grutowski v. Steamoat Lake Guides & Outfitters*, *Inc.,* 1998 WL 962042 (E.D. Pa., 1998); *Fix My PC, L.L.C. v. N.F.N. Associates, Inc.,* 48 F.Supp.2d 640 (N.D. Tex., 1999); *Origin Instruments Corp. v. Adaptive Computer Systems, Inc.,* 1999 WL 76794 (N.D. Tex., 1999); *Telephone Audio Productions, Inc. v. Smith*, 1998 WL

December 1997 Ninth Circuit decision, the Court was asked to determine whether the use of an allegedly infringing service mark on a Web site was sufficient grounds for asserting personal jurisdiction.[67] Both Cybersell Arizona, the owner of the "Cybersell" federal service mark, and Cybersell Florida provided Internet marketing and consulting services.[68] Cybersell Florida's presence in Arizona was limited to a Web site advertising its services and inviting interested parties to contact it for additional information. The Court's analysis followed the *Zippo* approach by attempting to ascertain the nature and quality of Cybersell Florida's Web based activity. As part of its analysis, the Court considered the passive nature of the site, the fact that no Arizonian other than Cybersell Arizona actually visited the site, and the fact that there was no evidence that an Arizonian entered into any contractual relationships with Cybersell Florida.[69] Considering these factors and its approval of the *Zippo's* summation of the law, the Court concluded that it could not properly assert jurisdiction in this matter.

Similarly, in *Mink v. AAAA Development LLC*,[70] the plaintiff, a computer program developer, filed suit in Texas district court against a Vermont corporation who allegedly conspired to copy the plaintiff's computer program. The district court refused to assume personal jurisdiction in the case and dismissed the suit. On appeal to the 5th Circuit, the plaintiff argued that Texas was the proper forum since the defendant corporation's Web site was accessible from that state. The court dismissed the appeal on the grounds that while the Web site provided users with a printable mail-in order form, an e-mail address, a toll-free telephone number and a mailing address, the fact that no orders were taken through the site meant that it was nothing more than a passive advertisement.

---

159932 (N.D. Tex., 1998); *Thompson v. Handa-Lopez, Inc.*, 998 F. Supp. 738 (W.D. Tex., 1998); *America Online, Inc. v. Huang*, 106 F.Supp.2d 848 (E.D. Va., 2000); *TELCO Communications v. An Apple A Day*, 977 F.Supp.404 (E.D. Va., 1997); *Conseco, Inc. v. Hickerson*, 698 N.E. 2d 816 (Ind.App., 1998); *State by Humphrey v. Granite Gate Resorts, Inc.,* (Minn. App., 1997).
[67] *Cybersell, Inc. v. Cybersell, Inc.,* 130 F. 3d. 414 (9th Cir. 1997).
[68] Interestingly, the principals behind Cybersell Arizona were Laurence Canter and Martha Siegel, attorneys who are infamous among Web users as the first Internet "spammers" or junk emailers.
[69] 130 F. 3d at 415.
[70] 190 F.3d 333 (5th Cir. (Tex.), 1999)

In *GTE New Media Services Incorporated v. Ameritech Corporation, et al.*,[71] the court was asked to assert jurisdiction over a company that was providing national Yellow Pages directory services over the Internet. Applying the passive versus active test, the court noted that the defendants maintained an interactive Web site that connected them with the District of Columbia. This fact, coupled with the fact that the defendants derived substantial advertising revenues from the directory sites when D.C. residents accessed and utilized its Internet Yellow Pages, led the court to exercise personal jurisdiction.

Finally, in *Desktop Technologies, Inc. v. Colorworks Reproduction & Design, Inc.*,[72] the plaintiff, a Pennsylvania corporation, filed suit in Pennsylvania district court against a Canadian company that carried on business exclusively in western Canada. The complaint alleged trademark infringement and breach of state unfair competition law based on the defendant's use of a trademark as its domain name on the Internet. The defendant brought a motion to dismiss the action for lack of personal jurisdiction. Citing the fact that its Internet presence and e-mail links were its only contacts with Pennsylvania, it noted that had never entered into any contracts in Pennsylvania nor sold anything in the state. Applying *Zippo* to these facts, the court ruled that the level of interactivity available on the defendant corporation's Web site did not justify exercising specific personal jurisdiction over the defendant, which was not doing business over the Internet with Pennsylvania residents.

Canadian courts signaled their approval of the *Zippo* approach in *Braintech Inc. v. Kostiuk*.[73] This 1999 British Columbia Court of Appeal case, the first Canadian appellate level decision to address the Internet jurisdiction issue, involved a series of allegedly defamatory messages posted on a stock chat site by a B.C. resident. Braintech, a B.C. based company, sued the poster in a Texas court, which awarded the company roughly $400,000 in damages.

---

[71] 44 F.Supp.2d 313 (D.D.C., 1999)
[72] 1999 WL 98572 (E.D. Pa., 1999)
[73] *Braintech, supra.*

When the company returned to B.C. to enforce the judgment, the B.C. courts examined the appropriateness of the Texas court's assertion of jurisdiction over the dispute. Adopting the passive versus active test by citing directly from the *Zippo* case, the B.C. Court of Appeal ruled that the Texas court had improperly asserted its jurisdiction. It argued that the postings were passive in nature and thus provided insufficient grounds to grant the Texas court authority over the case. Braintech's appeal to the Canadian Supreme Court was denied in early March 2000.[74]

The widespread approval for the *Zippo* test should come as little surprise. The uncertainty created by the Internet jurisdiction issue led to a strong desire for a workable solution that provided a fair balance between the fear of a lawless Internet and one burdened by over-regulation. The *Zippo* case seemed the best available alternative, particularly in light of the *Inset* line of cases which illustrated that the alternative might well be the application of jurisdiction by any court, anywhere. The court in *Neato v. Stomp LLC*, a 1999 federal court case in California, aptly summarized the competing policy positions by noting:

> The Court also recognizes that such a broad exercise of personal jurisdiction over defendants who engage in commerce over the Internet might have devastating effects on local merchants and small businesses that seek to expand through the Internet. These small businesses make up the backbone of the American economy and should not have to bear the burden of defending suits in distant fora when they mean only to allow local consumers to buy their wares from the convenience of their own homes. This concern must be balanced against the ability of a distant consumer to press its cause against a defendant who uses the Internet to do business within the forum while remaining outside the boundaries of the jurisdiction. Although a Plaintiff who seeks relief from the courts must be willing to overcome many of the hurdles that the litigation process imposes, it is the merchants who seek to sell their goods only to consumers in a particular geographic that can control the location of resulting lawsuits…. when a merchant seeks the benefit of engaging in unlimited interstate commerce over the Internet, it runs the risk of being subject to the process of the courts of those states.[75]

It must also be remembered that the *Zippo* passive versus active test was grounded in traditional jurisdictional principles. The analysis conducted as part of the test draws heavily from a foreseeability perspective, which suggests that it is not foreseeable for the owner of a passive Web site to face the prospect of getting hauled into court in multiple jurisdictions worldwide.  Conversely, as the Court in *Neato* recognized, the active e-commerce enabled Web site owner must surely foresee the possibility of disputes arising in other jurisdictions and recognize that those courts are entitled to protect local residents by applying local law and asserting jurisdiction.

---

[74] *Supreme Court of Canada - Judgments in Leave Applications*, Supreme Court of Canada (9 March 2000), online: Lexum <http://www.lexum.umontreal.ca/csc-scc/en/com/2000/html/00-03-09.3a.html> (date accessed: 31 March 2001)

[75] *Stomp Inc. v. Neato LLC*, 61 F. Supp. 2d 1074, 1080-1 (C.D. Cal. 1999). [hereinafter *Neato*]

Most importantly, in an emphatic repudiation of the "Internet as a separate jurisdiction" approach, the *Zippo* case made it explicit that local law still applies to the Internet.  Although it may at times be difficult to discern precisely whose law applies, there is little doubt that at least one jurisdiction, if not more, can credibly claim jurisdiction over any given Internet dispute.  With this principle in hand, the *Zippo* Court sent a clear signal to the Internet community that courts were willing to establish a balanced approach to Internet jurisdiction.

B.      The Shift Away from *Zippo*

Despite the widespread acceptance of the *Zippo* doctrine (and indeed the export of the test to other countries including Canada), cracks in the test began to appear late in 1999.  In fact, closer examination of the case law indicates that by 2001, many courts were no longer strictly applying the *Zippo* standard but rather were using other criteria to determine when assertion of jurisdiction was appropriate.[76]

Numerous judgments reflect that courts in the U.S. moved toward a broader, effects-based approach when deciding whether or not to assert jurisdiction in the Internet context. Under this new approach, rather than examining the specific characteristics of a

---

[76] In addition to the cases discussed infra, see also *Search Force v. DataForce Intern.*, 112 F.Supp.2d 771 (S.D. Ind., 2000); *Neato v. Great Gizmos*, 2000 WL 305949 (D. Conn., 2000); *Bochan v. La Fontaine*, 68 F.Supp.2d 701 (E.D. Va., 1999*); Rothschild Berry Farm v. Serendipity Group LLC*, 84 F.Supp.2d 904 (S.D. Ohio, 1999); *Uncle Sam's Safari Outfitters, Inc. v. Uncle Sam's Navy Outfitters – Manhattan, Inc.*, 96 F.Supp.2d 919 (E.D. Mo., 2000), *Compuserve v. Patterson*, 89 F.3d 1257 (6th Cir. 1996); *Neogen Corp. v. Neo Gen Screening, Inc.*, 109 F.Supp.2d 724 (W.D. Mich., 2000); *Panavision Intern., L.P. v. Toeppen*, 141 F.3d 1316 (9th Cir. (Cal), 1998).

Web site and its potential impact, courts focused their analysis on the actual effects that the Web site had in the jurisdiction. Indeed, courts are now relying increasingly on the effects doctrine that was established by the U.S. Supreme Court in *Calder v. Jones*.[77]

This doctrine holds that personal jurisdiction over a defendant is proper when a) the defendant's intentional tortious actions b) expressly aimed at the forum state c) causes harm to the plaintiff in the forum state, of which the defendant knows is likely to be suffered. In *Calder*, a California entertainer sued a Florida publisher for libel in a California district court. In ruling that personal jurisdiction was properly asserted, the Court focused on the effects of the defendant's actions. Reasoning that the plaintiff lived and worked in California, spent most of her career in California, suffered injury to her professional reputation in California, and suffered emotional distress in California, the Court concluded that defendant had intentionally targeted a California resident and thus it was proper to sue the publisher in that state.

The application of the Calder test can be clearly seen in an Internet context in *Blakey v. Continental Airlines, Inc.*,[78] an online defamation case involving an airline employee, living in Seattle and based out of Houston. The employee filed suit in New Jersey against her co-employees, alleging that they published defamatory statements on employer's electronic bulletin board, and against her employer, a New Jersey-based corporation, alleging that it was liable for the hostile work environment arising from the statements. The lower court granted the co-employees' motion to dismiss for lack of personal jurisdiction and entered summary judgment for the employer on the hostile work environment claim.

In reversing the ruling, the New Jersey Supreme Court found that defendants who published defamatory electronic messages with the knowledge that the messages would be published in New Jersey could properly be held subject to the state's jurisdiction. The court applied the effects doctrine and held that while the actions causing the effects in

---

[77] 465 U.S. 783 (1984)
[78] 164 N.J. 38 (N.J. 2000)

New Jersey were performed outside the state, this did not prevent the court from asserting jurisdiction over a cause of action arising out of those effects.

The broader effects-based analysis can also be seen moving beyond the defamatory tort action at issue in *Calder* and *Blakey* to a range of disputes including intellectual property and commercial activities. On the intellectual property front, *Nissan Motor Co. Ltd. v. Nissan Computer Corporation,*[79] typifies the approach. The plaintiff, an automobile manufacturer, filed a complaint in California district court against a Massachusetts-based computer seller, alleging that the defendant's "nissan.com" and "nissan.net" Internet domain names infringed on its "Nissan" trademark. Prompting the complaint was an allegation that the defendant altered the content of its "nissan.com" Web site to include a logo that was similar to the plaintiff's logo, as well as include links to automobile merchandisers and auto-related portions of search engines. In October 1999 the parties met to discuss the possibility of transferring the nissan.com domain name. These negotiations were ultimately unsuccessful. The defendant brought a motion to dismiss for lack of personal jurisdiction and improper venue and the plaintiff brought a motion for a preliminary injunction in March 2000.

In considering the defendant's motion, the court relied on the effects doctrine to assert jurisdiction, ruling that the defendant had intentionally changed the content of its Web site to exploit the plaintiffs' goodwill and to profit from consumer confusion. Moreover, since the plaintiff was based in California, the majority of the harm was suffered in the forum state. The court rejected the defendant's argument that it was not subject to personal jurisdiction because it merely operated a passive Web site. Although the defendant did not sell anything over the Internet, it derived advertising revenue through the intentional exploitation of consumer confusion. This fact, according to the court, satisfied the *Cybersell* requirement of "something more" in that it established that the defendant's conduct was deliberately and substantially directed toward the forum state.

---

[79] 89 F.Supp.2d 1154 (C.D. Cal., 2000)

Similarly, in *Euromarket Designs Inc. v. Crate & Barrel Ltd.*,[80] the issue before the court was whether an Illinois-based company could sue an Irish retailer with an interactive Web site that allowed Illinois residents to order goods for shipment to a foreign address in a local court for trademark infringement. The court noted that the pivotal considerations in resolving this issue were whether the defendant purposefully and deliberately availed itself of the forum and whether the defendant's conduct and connection with the forum was such that it should reasonably anticipate being hauled into court there. The court stated that the defendant deliberately established minimum contacts with Illinois and purposefully availed itself of the privilege of conducting activities in Illinois under the effects doctrine set out in *Calder*.

The court concluded that the defendant's actions established jurisdiction under the effects doctrine since: a) if plaintiff's trademark was infringed, the injury would be felt primarily in Illinois; b) the defendant intentionally and purposefully directed its actions toward Illinois and the plaintiff, an Illinois corporation, allegedly causing harm to the plaintiff in Illinois; and c) the defendant knew that harm would likely be suffered in Illinois.[81]

Courts have also refused to assert jurisdiction in a number cases based on what is best described as insufficient commercial effects. For example, in *People Solutions, Inc. v. People Solutions, Inc.*[82] the defendant, a California-based corporation, moved to dismiss a trademark infringement suit brought against it by a Texas-based corporation of the same name. The plaintiff argued that the suit was properly brought in Texas since the defendant owned a Web site that could be accessed and viewed by Texas residents. The site featured several interactive pages that allowed customers to take and score performance tests, download product demos, and order products online.

The court characterized the site as interactive but refused to assert jurisdiction over the matter. Relying on evidence that no Texans had actually purchased from the

---

[80] 96 F.Supp.2d 824 (N.D. Ill., 2000)
[81] *Id.* at 836.
[82] 2000 WL 1030619 (N.D. Tex., 2000).

Web site, the court held that "[p]ersonal jurisdiction should not be premised on the mere possibility, with nothing more, that Defendant may be able to do business with Texans over its Web site."[83] Instead, the plaintiff had to show that the defendant had "purposefully availed itself of the benefits of the forum state and its laws." [84]

A copyright dispute over craft patterns yielded a similar result in *Winfield Collection, Ltd. v. McCauley*.[85] The plaintiff, a Michigan-based manufacturer of craft patterns, filed a complaint in Michigan district court accusing the defendant, a resident of Texas, of infringing copyrighted craft patterns that it had supplied to the defendant. The defendant moved to dismiss the suit for lack of personal jurisdiction. The plaintiff argued that the court could exercise personal jurisdiction because a) the defendant had sold crafts made with the plaintiff's patterns to Michigan residents on two occasions, and b) the defendant maintained an interactive Web site that could send and receive messages.

The court refused to assert jurisdiction, dismissing both arguments. With respect to the plaintiff's first argument, the court focused on the fact that the sales were in fact concluded on eBay, an online auction site. Since the items were sold to the highest bidder, the defendant had no advance knowledge about where the products would be sold. As such, she did not purposefully avail herself of the privilege of doing business in Michigan.

In response to the plaintiff's second argument, the court held that it was not prepared to broadly hold "that the mere act of maintaining a Web site that includes interactive features *ipso facto* establishes personal jurisdiction over the sponsor of that Web site anywhere in the United States."[86] In its judgment the court noted that the plaintiff had provided it with the unpublished opinion in a case called *Amway v. Proctor & Gamble*. In that case, the court held that "something more" than mere activity should be required to assert personal jurisdiction and found that "something more" to be the effects doctrine. The court held that the plaintiff could not rely on that doctrine since it failed to identify a continuing relationship with Michigan or with any resident of Michigan.

---

[83] 2000 WL 1030619 at 5.
[84] *Id*.
[85] 105 F.Supp.2d 746 (E.D. Mich., 2000).
[86] *Id*. at 751.

One of the strongest criticisms of the *Zippo* doctrine can be found in *Millenium Enterprises, Inc. v. Millenium Music L.P.*,[87] another case in which the court found insufficient commercial effects and therefore declined to assert jurisdiction. The defendant, a South Carolina corporation, sold products both offline and on the Web. The plaintiffs, an Oregon-based corporation, sued the defendants in Oregon district court for trademark infringement. The defendant filed a motion to dismiss for lack of personal jurisdiction. After canvassing numerous Internet jurisdiction cases decided by the Ninth Circuit, as well as *Zippo*, the court stated:

> [T]he middle interactive category of Internet contacts as described in *Zippo* needs further refinement to include the fundamental requirement of personal jurisdiction: "deliberate action" within the forum state in the form of transactions between the defendant and residents of the forum or conduct of the defendant purposefully directed at residents of the forum state. See Calder, 465 U.S at 798-90. This, in the court's view, is the "something more" that the Ninth Circuit intended in Cybersell and Panavision.[88]

Applying this reasoning to the facts before it, the court allowed the defendants' motion to dismiss the case on the grounds that the defendants had consummated no transaction and had not made deliberate and repeated contacts with Oregon through their Web site, such that they could reasonably anticipate being hauled into Oregon court. In its concluding remarks, the court said:

> For all of these reasons, this court will not abandon the basic principle that defendants must have taken some action to direct their activities in the forum so as to "purposely avail" themselves of the privilege of doing business within Oregon. The timeless and fundamental bedrock of personal jurisdiction assures us all that a defendant will not be "haled" into a court of a foreign jurisdiction based on nothing more than the foreseeability or potentiality of commercial activity with the forum state....[D]efendants cannot reasonably anticipate that they will be brought before this court, simply because they advertise their products

---

[87] 33 F.Supp.2d 907 (D. Or., 1999)
[88] *Id*. at 921

through a global medium which provides the capability of engaging in commercial transactions.[89]

Although the case law illustrates that there was no single reason for the courts to shift away from the *Zippo* test, a number of themes do emerge. First, the test simply doesn't work particularly well in every instance. For example, with courts characterizing chat room postings as passive in nature,[90] many might be inclined to dismiss cases involving allegedly defamatory or harassing speech on jurisdictional grounds. Such speech may often be targeted toward a particular individual or entity located in a jurisdiction different from that of the poster or the chat site itself. Characterizing this act as passive does not result in a desirable outcome since the poster knows or ought to know that the effect of their posting will be felt most acutely in the home jurisdiction of the target. If the target is unable to sue locally due to a strict adherence to the passive versus active test, the law might be seen as encouraging online defamatory speech by creating a jurisdictional hurdle to launching a legal claim.

The *Zippo* test also falls short when active sites are at issue, as the court in *People Solutions* recognized. That Court's request for evidence of actual sales within the jurisdiction illustrates that the mere potential to sell within a jurisdiction does not necessarily make a Web site active. While the active Web site may want to sell into every jurisdiction, the foreseeability of a legal action is confined primarily to those places where actual sales occur. The *Zippo* test does not distinguish between actual and potential sales, however, but rather provides that the mere existence of an active site is sufficient to assert jurisdiction.

The problem with the *Zippo* test is not limited to inconsistent and often undesirable outcomes. The test also encourages a perverse behaviour that runs contrary to public policy related to the Internet and e-commerce. Most countries have embraced the potential of e-commerce and adopted policies designed to encourage the use of the

---

[89] *Id.* at 923.
[90] In Canada, *see*, *Braintech*, *supra*; in the U.S., *see*, *Barrett v. Catacombs Press*, 44 F. Supp. 2d 717 (E.D. Pa. 1999).

Internet for commercial purposes.[91] The *Zippo* test, however, inhibits e-commerce by effectively discouraging the adoption of interactive Web sites. Prospective Web site owners who are concerned about their exposure to legal liability will rationally shy away from developing active Web sites since such sites increase the likelihood of facing lawsuits in far-off jurisdictions. Instead, the test encourages passive Web sites that feature limited legal exposure and therefore present limited risk. Since public policy aims are to increase interactivity and the adoption of e-commerce (and in doing so, enhance consumer choice and open new markets for small and medium sized businesses), the *Zippo* test acts as a barrier to that policy approach.

One of the primary reasons for the early widespread support for the *Zippo* test was the desire for increased legal certainty for Internet jurisdiction issues. While the test may not have been perfect, supporters felt it offered a clear standard that would allow businesses to conduct effective legal risk analysis and make rational choices with regard to their approach to the Internet.[92]

---

[91] The Canadian government's e-commerce policy states the following:
> "On September 22, 1998, the Prime Minister announced Canada's Electronic Commerce Strategy, outlining initiatives designed to establish Canada as a world leader in the adoption and use of electronic commerce. Working in close collaboration with the private sector, the federal government has concentrated on creating the most favourable environment possible in areas which are critical to the rapid development of e-commerce."

*Electronic Commerce in Canada - Canadian Strategy*, Industry Canada - Task Force on Electronic Commerce Task Force (IC-TFEC), online: IC-TFEC <http://www.ecom.ic.gc.ca/english/60.html> (date accessed: 30 March 2001).

Similarly, the U.S. government's e-commerce policy states the following:
> "Commerce on the Internet could total tens of billions of dollars by the turn of the century. For this potential to be realized fully, governments must adopt a non-regulatory, market-oriented approach to electronic commerce, one that facilitates the emergence of a transparent and predictable legal environment to support global business and commerce. Official decision makers must respect the unique nature of the medium and recognize that widespread competition and increased consumer choice should be the defining features of the new digital marketplace."

*United States Government Electronic Commerce Policy - A Framework for Global Electronic Commerce*, The White House (1 July 1997), online: United States Government Electronic Commerce Policy <http://www.ecommerce.gov/framewrk.htm.> (date accessed: 30 March 2001).

[92] Professor John Gedid noted the following at an international conference on Internet jurisdiction:

In the final analysis, however the *Zippo* test simply does not deliver the desired effect. First, the majority of Web sites are neither entirely passive nor completely active. Accordingly, they fall into the "middle zone" which requires courts to gauge all relevant evidence and determine whether the site is "more passive" or "more active." With many sites falling into this middle zone, their legal advisors are frequently unable to provide a firm opinion on how any given court might judge the interactivity of the Web site.

Second, distinguishing between passive and active sites is complicated by the fact that some sites may not be quite what they seem. For example, sites that feature content best characterized as passive, may actually be using cookies or other data collection technologies behind the scenes unbeknownst to the individual user.[93] Given the value accorded to personal data,[94] its collection is properly characterized as active, regardless of whether it occurs transparently or surreptitiously.

---

"The Zippo opinion is comprehensive, thorough and persuasive. In it the court displays substantial understanding, or at least willingness to learn about, operations on the Internet; comprehension of the idea that the principles of International Shoe control this new technology; and a willingness to reason analogically to apply the International Shoe criteria to the problem of cyberspace jurisdiction. The court's review of precedents is sweeping and thorough, and its logic is compelling. The Zippo court fully understood and explained difficult precedents, so that they could be understood in terms of the International Shoe criteria. While there are some who would question the approach on the theories that it does not go far enough or that it goes too far, nevertheless, it is an attempt at stating a more comprehensive and coherent approach Internet jurisdiction cases. The result was that the Zippo opinion is probably the most persuasive and influential opinion that has been published on the subject of cyberspace jurisdiction. So many subsequent decisions have drawn upon and applied the Zippo analysis that in the few years since it appeared it has become the leading authority on cyberspace jurisdiction."

Professor John L. Gedid, *Minimum Contacts Analysis In Cyberspace--Sale Of Goods And Services (DRAFT-WORK IN PROGRESS),* Internet Law and Policy Forum - Jurisdiction: Building Confidence in a Borderless Medium (Montréal, Canada, 26-27 July 1999), online: ILPF <http://ilpf.org/confer/present99/gedid_addl.htm> (date accessed: 2 April 2001). See, also, Charles H. Fleischer ,*Will The Internet Abrogate Territorial Limits On Personal Jurisdiction*?, 33 Tort & Ins. L.J. 107 (1997); Michael J. Sikora III, *Beam Me Into Your Jurisdiction: Establishing Personal Jurisdiction Via Electronic Contacts In Light Of The Sixth Circuit's Decision In Compuserve,Inc. v. Patterson*, 27 Cap. U. L. Rev. 163, 184-5 (1998).

[93] Jerry Kang, *Information Privacy in Cyberspace Transactions*, 50 Stan. L. Rev. 1193, 1226-9 (1998).

[94] *Ibid*.

Third, it is important to note that the standards for what constitutes an active or passive Web site is constantly shifting. When the test was developed in 1997, an active Web site might have featured little more than an email link and some basic correspondence functionality. Today, sites with that level of interactivity would likely be viewed as passive, since the entire spectrum of passive versus active has shifted upward together with improved technology. In fact, it can be credibly argued that sites must constantly re-evaluate their position on the passive versus active spectrum as Web technology changes.

Fourth, the effectiveness of the *Zippo* test is no better even if the standards for passive and active sites remain constant. With the expense to create a sophisticated Web site now easily in excess of $100,000,[95] few organizations will invest without anticipating some earning potential for their Web-based venture. Since revenue is typically the hallmark of active Web sites, most new sites are likely to feature interactivity and be categorized as active sites. From a jurisdictional perspective, this produces an effect similar to that found in the *Inset* line of cases – any court anywhere can assert jurisdiction over a Web site since virtually all sites will meet the *Zippo* active benchmark.

In light of the ever-changing technological environment and the shift toward predominantly active Web sites, the effectiveness of the *Zippo* doctrine is severely undermined regardless of how it develops. If the test evolves with the changing technological environment, it fails to provide much needed legal certainty. If the test remains static to provide increased legal certainty, it risks becoming irrelevant as the majority of Web sites meet the active test standard.

Part IV – Toward a Trio of Targets

---

[95] David Legard, *Average Cost to Build E-commerce Site: $1 Million*, The Standard (31 May 1999), online: The Standard <http://www.thestandard.com/article/article_print/0,1153,4731,00.html> (date accessed: 31 March 2001).

Given the inadequacies of the *Zippo* passive versus active test, it is now fitting to identify a more effective standard for determining when it is appropriate to assert jurisdiction in cases involving predominantly Internet-based contacts. With the benefit of the *Zippo* experience, the new test should remain technology neutral so as to a) remain relevant despite ever-changing Web technologies, b) create incentives that, at a minimum, do not discourage online interactivity, and c) provide sufficient certainty so that the legal risk of operating online can be effectively assessed in advance.

The solution submitted here is to move toward a targeting-based analysis. Unlike the *Zippo* approach, a targeting analysis would seek to identify the intentions of the parties and to assess the steps taken to either enter or avoid a particular jurisdiction. Targeting would also lessen the reliance on effects-based analysis, the source of considerable uncertainty since Internet-based activity can ordinarily be said to create some effects in most jurisdictions.

A targeting approach is not a novel idea. Several U.S. courts have factored targeting considerations into their analysis of the appropriateness of asserting jurisdiction over Internet-based activities. For example, in *Bancroft & Masters, Inc. v. Augusta National Inc.*, a dispute over the masters.com domain name, the 9th Circuit Court of Appeal noted:

> To meet the effects test, the defendant must have (1) committed an intentional act, which was (2) expressly aimed at the forum state, and (3) caused harm, the brunt of which is suffered and which the defendant knows is likely to be suffered in the forum state. See Panavision Int'l, L.P. v. Toeppen, 141 F.3d 1316, 1321 (9th Cir.1998). Subsequent cases have struggled somewhat with Calder 's import, recognizing that the case cannot stand for the broad proposition that a foreign act with foreseeable effects in the forum state always gives rise to specific jurisdiction. We have said that there must be "something more," but have not spelled out what that something more must be. See Panavision, 141 F.3d at 1322. We now conclude that "something more" is what the Supreme Court described as "express aiming" at the forum state. See Calder, 465 U.S. at 789, 104 S.Ct. 1482. *Express aiming is a concept that in the jurisdictional context hardly defines itself. From the available cases, we deduce that the requirement is satisfied when the defendant is alleged to have engaged in wrongful conduct targeted at a plaintiff whom the defendant knows to be a resident of the forum state.*[96] [emphasis added]

---

[96] Bancroft & Masters Inc. v. Augusta National Inc., 223 F.3d 1082, 1087 (9th Cir. 2000).

Targeting has also been raised in the context of online gambling cases, where U.S. courts have aggressively characterized offshore gambling sites as "targeting" local residents. In *People v. World Interactive Gaming*,[97] the court considered the targeting issue and stated:

> Wide range implications would arise if this Court adopted respondents' argument that activities or transactions which may be targeted at New York residents are beyond the state's jurisdiction. Not only would such an approach severely undermine this state's deep-rooted policy against unauthorized gambling, it also would immunize from liability anyone who engages in any activity over the Internet which is otherwise illegal in this state. A computer server cannot be permitted to function as a shield against liability, particularly in this case where respondents actively targeted New York as the location where they conducted many of their allegedly illegal activities.[98]

Targeting-based analysis has also become increasingly prevalent among international organizations seeking to develop global minimum legal standards for e-commerce. The OECD Consumer Protection Guidelines refer to the concept of targeting, stating that "business should take into account the global nature of electronic commerce and, wherever possible, should consider various regulatory characteristics of the markets they target."[99]

Similarly, the most recent draft of the Hague Conference on Private International Law's Draft Convention on Jurisdiction and Foreign Judgments includes provisions related to targeting.[100] During negotiations over the e-commerce implications of the draft convention in Ottawa in February 2001, delegates focused on targeting as a means of distinguishing when consumers should be entitled to sue in their home jurisdiction.

---

[97] 714 N.Y.S. 2d 844 (N.Y.Sup. 1999), 1999 N.Y. Misc. LEXIS 425 (S.C. N.Y. 1999).[hereinafter *World Interactive Gaming*].
[98] *Id*, at para. 9.
[99] OECD, *Recommendation of the OECD Council Concerning Guidelines for Consumer Protection in the Context of Electronic Commerce* (Paris, 9 December 1999), online: OECD <http://www.oecd.org/dsti/sti/it/consumer/prod/CPGuidelines_final.pdf. (date accessed: 31 March 2001)
[100] *Hague Conference On Private International Law: Preliminary Draft Convention on Jurisdiction and Foreign Judgments In Civil and Commercial Matters,* 30 October 1999, online: Hague Conference <http://www.hcch.net/e/conventions/draft36e.html> (date accessed: 31 March 2001).

Version 0.4a of Article 7 (3)(b) includes a provision which states that "activity by the business shall not be regarded as being directed to a State if the business demonstrates that it took reasonable steps to avoid concluding contracts with consumers habitually resident in that State."[101]

Targeting also forms the central consideration for securities regulators assessing online activity. As the U.S. Securities and Exchange stated in its release on the regulation of Internet-based offerings:

> The regulation of offers is a fundamental element of federal and some U.S. state securities regulatory schemes. Absent the transaction of business in the United States or with U.S. persons, however, our interest in regulating solicitation activity is less compelling. We believe that our investor protection concerns are best addressed through the implementation by issuers and financial service providers of precautionary measures that are reasonably designed to ensure that offshore Internet offers are not targeted to persons in the United States or to U.S. persons.[102]

The same targeting approach has met with approval in Canada,[103] the United Kingdom,[104] and other parts of the world.[105] In Canada, the Canadian Securities Association has adopted a policy that requires online securities offerings to specifically exclude Canada in order to avoid the jurisdictional reach of Canadian securities regulators. According to the CSA, excluding Canada requires the use of a prominent disclaimer as well as reasonable precautions to ensure that securities are not sold to anyone in Canada.[106]

---

[101] *Id*, Article 7, Version 0.4a.
[102] *Interpretation: Re: Use of Internet Web Sites to Offer Securities, Solicit Securities Transactions, or Advertise Investment Services Offshore*, Release No. 33-17516, (March 23, 1998).
[103] *National Policy 47-210: Trading in Securities Using the Internet and Other Electronic Means*, (Notice NIN#98/72), online: British Columbia Securities Commission <http://www.bcsc.bc.ca/Policy/Nin98-72.pdf> (date accessed: 30 March 2001). [hereinafter Canadian Securities].
[104] Financial Services Authority, *Discussion Paper - The FSA's Approach To Regulation Of The Market Infrastructure*, (January 2000), online: FSA <http://www.fsa.gov.uk/pubs/discussion/d02.pdf> (date accessed: 31 March 2001).
[105] *IOSCO, supra.*
[106] *Canadian Securities, supra.*

The American Bar Association Internet Jurisdiction Project, a global study on Internet jurisdiction released in 2000, also recommended targeting as one method of addressing the Internet jurisdiction issue.[107] The report noted that:

> Today, entities seeking a relationship with residents of a foreign forum need not themselves maintain a physical presence in the forum. A forum can be "targeted" by those outside it and desirous of benefiting from a connecting with it via the Internet (assuming, of course, that the foreign actor is willing to confine its target to those with access to technology, a growing but still not universal subset of any forum's population). Such a chosen relationship will subject the foreign actor to both personal and prescriptive jurisdiction, so a clear understanding of what constitutes targeting is critical.[108]

It is the ABA's last point – that a clear understanding of what constitutes targeting is essential -- that requires careful examination and discussion. Without universally applicable standards for targeting assessment in the online environment, a targeting-based test is likely to leave further uncertainty in its wake. For example, the ABA's report refers to language as a potentially important determinant for targeting purposes. That criteria overlooks the fact that the development of new language translation capabilities may soon enable a Web site owner to display their site in the language of their choice, safe in the knowledge that visitors around the world will read the content in their own language through the aid of translation technologies.[109]

Targeting as the litmus test for Internet jurisdiction is only the first step in the development of a consistent test that provides increased legal certainty. The second, more challenging step is to identify the criteria to be used in assessing whether a Web site has indeed targeted a particular jurisdiction. This step is challenging because the criteria must meet at least two important standards. First, the criteria must be technology neutral so that the test remains relevant even as new technologies emerge. This would seem to disqualify criteria such as a Web site language or currency, which is susceptible to real-time conversion by newly emerging technologies.

---

[107] *Achieving Legal and Business Order in Cyberspace: A Report on Global Jurisdiction Issues Created By the Internet*, ABA, 2000. In the interests of full disclosure, it should be noted that the author was chair of the Sale of Services Working Group, one of nine working groups tasked with developing Internet jurisdiction recommendations.
[108] *Id*.
[109] Currently in beta, Google offers searchers the ability to configure their Google searching to translate automatically any results that appear in a foreign language. *See* <http://www.google.com/machine_translation.html> (date accessed: 3 April 2001).

Second, the criteria must be content neutral so that there is no apparent bias in favour of any single interest group or constituency. Internet jurisdiction is a particularly sensitive issue with several business groups lobbying for a "rule of origin" approach under which jurisdiction would always rest with the jurisdiction of the seller.[110] Consumer groups, meanwhile, have lobbied for a "rule of destination" approach that ensures that consumers can always sue in their home jurisdiction.[111] The origin versus destination debate has polarized both groups, making it difficult to reach a compromise that recognizes that effective consumer protection does not depend solely on which law applies, and acknowledge, as the *Neato* court does, that business must shoulder some of the risk arising from e-commerce transactions.[112]

To identify the appropriate criteria for a targeting test, we must ultimately return to the core jurisdictional principle – foreseeability. Foreseeability should not be based on a passive versus active Web site matrix, however. Rather, an effective targeting test requires an assessment of whether the targeting of a specific jurisdiction was itself foreseeable. Foreseeability in that context depends on three factors -- contracts, technology, and actual or implied knowledge. Forum selection clauses found in Web site terms of use agreements or transactional clickwrap agreements allow parties to mutually determine an appropriate jurisdiction in advance of a dispute. It therefore provides important evidence as to the foreseeability of being hauled into the courts of a particular jurisdiction. Newly-emerging technologies that identify geographic location constitute the second factor. These technologies, which challenge widely held perceptions about the Internet's architecture, may allow sites to target their content by engaging in "jurisdictional avoidance." The third factor, actual or implied knowledge, is a catch-all that incorporates targeting knowledge gained through the geographic location of tort victims, offline order fulfillment, financial intermediary records, and Web traffic.

---

[110] See, e.g., Global Business Dialogue on Electronic Commerce <http://www.gbde.org> (date accessed: 31 March 2001).
[111] See, e.g., Consumers International <http://www.consumersinternational.org> (date accessed: 31 March 2001).
[112] *Neato*, *supra*.

Although all three factors are important, no single factor should be determinative. Rather, each must be analyzed to adequately assess whether the parties have fairly negotiated a governing jurisdiction clause at a private contract level, whether the parties employed any technological solutions to target their activities, and whether the parties knew, or ought to have known, where their online activities were occurring. While all three factors should be considered as part of a targeting analysis, the relative importance of each will vary. Moreover, in certain instances, some factors may not factor at all. For example, a defamation action is unlikely to involve a contractual element, though the evidence from the knowledge factor is likely to prove sufficient to identify the targeted jurisdiction.

It is important to also note that the targeting analysis will not determine exclusive jurisdiction, but rather identify whether a particular jurisdiction can be appropriately described as having been targeted. The test does not address which jurisdiction is the *most* appropriate as between those jurisdictions that meet the targeting threshold.

A.     Contracts

The first of the three factors for the recommended targeting test considers whether either party has used a contractual arrangement to specify which law should govern. Providing parties with the opportunity to limit their legal risk by addressing jurisdictional concerns in advance through contract can be the most efficient and cost-effective approach to dealing with the Internet jurisdiction issue.

The mere existence of a jurisdictional clause within a contract, however, should not, in and of itself, be determinative of the issue. In addition to considering the two other targeting factors, the weight accorded to an online contract should depend upon the method used to obtain assent and the reasonableness of the terms contained in the contract.

Courts in both Canada and the U.S. have upheld the *per se* enforceability of an online contract,[113] commonly referred to as a clickwrap agreement. These agreements typically involve clicking on an "I agree" icon to indicate assent to the agreement. Given their ubiquity, it should come as little surprise to find that courts have been anxious to confirm their enforceability. For example, in the 1999 Ontario case of *Rudder v. Microsoft Corp.*,[114] a case involving a dispute over the validity of a forum selection clause, the court noted:

> It is plain and obvious that there is no factual foundation for the plaintiffs' assertion that any term of the Membership Agreement was analogous to "fine print" in a written contract. What is equally clear is that the plaintiffs seek to avoid the consequences of specific terms of their agreement while at the same time seeking to have others enforced. Neither the form of this contract nor its manner of presentation to potential members are so aberrant as to lead to such an anomalous result. To give effect to the plaintiffs' argument would, rather than advancing the goal of "commercial certainty", to adopt the words of Huddart J.A. in Sarabia, move this type of electronic transaction into the realm of commercial absurdity. It would lead to chaos in the marketplace, render ineffectual electronic commerce and undermine the integrity of any agreement entered into through this medium.[115]

Courts in the U.S. have been similarly supportive of forum selection clauses found in clickwrap contracts. In the recent *Kilgallen v. Network Solutions, Inc.*,[116] the court was faced with a dispute over the re-registration of a domain name. The plaintiff claimed that Network Solutions, the defendant, was in breach of contract when it transferred its domain name to a third party. Network Solutions defended its actions by countering that the plaintiff had failed to make the annual payment necessary to maintain the domain. Moreover, it sought to dismiss the action on the grounds that the proper court was the Eastern District of Virginia, as specified in the registrant agreement. The federal court in Massachusetts agreed, ruling that forum selection clauses are enforceable unless proven unreasonable under the circumstances.

---

[113] *Rudder, supra*; *Killagen v. Network Solutions*, 99 F. Sup. 2d 125 (D. Mass. 2000); *Graves v. Pikulski*, 115 F. Supp. 2d 931 (S.D. Ill., 2000).
[114] *Rudder, supra.*
[115] *Rudder*, *supra* at para 16.
[116] *Kilgallen, supra.*

Notwithstanding the apparent support for enforcing forum selection clauses within clickwrap agreements, the presence of such a clause should only serve as the starting point for analysis. A court must first consider how assent to the contract was obtained. If the agreement is a standard clickwrap agreement in which the user was required to positively indicate their agreement by clicking on an "I agree" or similar icon, the court will likely deem this to be valid assent.

Many jurisdictional clauses are not found in a clickwrap agreement, however, but rather are contained in the terms of use agreement on the Web site. The terms typically provide that users of the Web site agree to all terms contained therein by virtue of their use of the Web site.

The validity of this form of contract, in which no positive assent is obtained and the Web site visitor is unlikely to have read the terms, stands on shakier ground. Two recent U.S. cases have considered this form of contract and have muddled the issue by delivering conflicting decisions. In *Ticketmaster v. Tickets.com*,[117] a dispute over links between rival event ticket sites, the court considered the enforceability of the terms and conditions page found on the Ticketmaster site and concluded:

> The motion to dismiss the second claim (breach of contract) is founded on the "terms and conditions" set forth on the home page of the Ticketmaster site. This provides that anyone going beyond the home page agrees to the terms and conditions set forth, which include that the information is for personal use only, may not be used for commercial purposes, and no deep linking to the site is permitted. In defending this claim, Ticketmaster makes reference to the "shrink-wrap license" cases, where the packing on the outside of the CD stated that opening the package constitutes adherence to the license agreement (restricting republication) contained therein. This has been held to be enforceable. That is not the same as this case because the "shrink-wrap license agreement" is open and obvious and in fact hard to miss. Many web sites make you click on "agree" to the terms and conditions before going on, but Ticketmaster does not. Further, the terms and conditions are set forth so that the customer needs to scroll down the home page to find and read them. Many customers instead are likely to proceed to the event page of interest

---

[117] 2000 WL 525390 (C.D.Cal.,2000), 2000 U.S. Dist. LEXIS 4553 (C.D. Cal. 2000).

rather than reading the "small print." *It cannot be said that merely putting the terms and conditions in this fashion necessarily creates a contract with any one using the web site*.[118] [emphasis added]

The Ticketmaster case suggests that mere inclusion of a forum selection or other jurisdictional clause, within the terms and conditions, may not be enforceable since the term is not brought sufficiently to the attention of the user.

Several months after the Ticketmaster decision, another federal court adopted a different approach. *Register.com, Inc. v. Verio, Inc*[119] involved a dispute over Verio's use of automated software to access and collect the domain name registrant contact information contained in the Register.com WHOIS database. Verio collected such data and then used it for marketing purposes. Register.com provided the following terms and conditions for those wishing to access its WHOIS database:

> By submitting a WHOIS query, you agree that you will use this data only for lawful purposes and that, under no circumstances will you use this data to: (1) allow, enable, or otherwise support the transmission of mass unsolicited, commercial advertising or solicitations via direct mail, electronic mail, or by telephone; or (2) enable high volume, automated, electronic processes that apply to Register.com (or its systems). The compilation, repackaging, dissemination or other use of this data is expressly prohibited without the prior written consent of Register.com. Register.com reserves the right to modify these terms at any time. By submitting this query, you agree to abide by these terms.[120]

Unlike the Ticketmaster case, the court in Register.com ruled that these terms were binding on users, despite the absence of a clear manifestation of assent.

While the form of assent may call into question the validity of an online contract, the actual terms of the contract itself is of even greater consequence. Courts are required to consider the reasonableness of the terms of a contract as part of their analysis. Within the context of a jurisdictional inquiry, several different scenarios may lead to the court discounting the importance of the contract as part of a targeting analysis.

---

[118] *Id*, at para. 3.
[119] 126 F.Supp.2d 238 (S.D.N.Y. 2000).
[120] *Id*, at ff para. 1.

A court may simply rule that the forum selection clause is unenforceable in light of the overall nature of the contract. This occurred in *Mendoza v. AOL*,[121] a recent California case involving a disputed ISP bill. After Mendoza sued AOL in California state court, AOL responded by seeking to have the case dismissed on the grounds that the AOL service contract contains a forum selection that requires all disputes arising from the contract to be brought in Virginia. The court surprised AOL by refusing to enforce the company's terms of service agreement on the grounds that "it would be unfair and unreasonable because the clause in question was not negotiated at arm's length, was contained in a standard form contract, and was not readily identifiable by plaintiff due the small text and location of the clause at the conclusion of the agreement."[122]

Though cases such as Mendoza are the exception rather than the rule, they do point to the fact that a forum selection clause will not always be enforced, particularly in consumer disputes where the provision may be viewed by a court as too onerous given the small amount at issue.[123]

Courts may also be unwilling to enforce such clauses where the court perceives that the party is seeking to contract out of the jurisdiction with the closest tie to the parties. Courts must be particularly vigilant in such cases to ensure that forum selection clauses are not used to create a "race to the bottom" effect whereby parties select jurisdictions with lax regulations in an attempt to avoid more onerous regulations in the home jurisdictions of either the seller or purchaser.[124] Aggressive courts may also be unwilling to enforce a clause with no tie to the jurisdiction. In *Standard Knitting, Ltd. v.*

---

[121] *Mendoza*, *supra*.
[122] *Ibid*.
[123] For another recent example, see, *Williams v. AOL*, (http://www.socialaw.com/superior/000962.html) in which a Massachusetts state court refused to enforce the AOL forum selection clause in a class action suit over AOL system software.
[124] For example, the Wall Street Journal reports that Bermuda has become a haven for dot-com operations seeking to avoid tax and other regulatory measures in North America. See, M. Allen, *As Dot-Coms Go Bust in the U.S., Bermuda Hosts a Little Boomlet*, Wall Street Journal (8 January 2001).

*Outside Design, Inc.*,[125] an Internet jurisdiction case involving a Canadian plaintiff, for example, the federal court in Pennsylvania transferred the case to Washington state after it found that venue to be more convenient for the parties.

An alternative to dictating jurisdiction terms to the consumer and risking a court's refusal to enforce, is to provide the consumer with the opportunity to self-declare their jurisdiction. The advantage with this approach is that the business can refuse to deal with the consumer if they self-declare a jurisdiction with increased legal risk. For example, Expedia, a leading online travel site, asks users to indicate their home jurisdiction prior to using the service.[126] If the user indicates the U.S. as their home jurisdiction, they remain at the Expedia.com site. If the user lists Canada as their home jurisdiction, they are transferred to Expedia.ca, a Canadian-specific site. If the user lists Mexico as their home jurisdiction, the site advises them that Expedia is unable to provide service at the present time due to regulatory constraints.

An additional advantage to this approach is that the business should be able to rely on the consumer's self-declaration. If the consumer intentionally proffers incorrect information – they reside in Mexico but declare that the U.S. is their home jurisdiction – Expedia should be able to rely on the consumer statement to ensure that they do not run afoul of Mexican regulatory law since they were clearly targeting their activity to the U.S.[127]

Despite the potential advantages of self-declaration, courts have ruled that companies cannot rely on the self-declaration of a user where they know or suspect it to be false. For example, in *People v. World Interactive Gaming*,[128] an online gambling case, the court rejected attempts by the online casino to limit registration to gamblers resident in a state that permits gambling. In particular the court noted that:

> In opening an account, users were asked to enter their permanent address. A user which submitted a permanent address in a state that permitted land-based gambling, such as Nevada, was granted permission to gamble. Although a user which entered a state such as New York, which does not permit land-based gambling, was denied permission to gamble, because the software does not verify the user's actual location, a user initially denied access, could easily circumvent the denial by changing the state entered to that of Nevada, while remaining physically in New York State. The user could then log onto the GCC casino and play virtual slots, blackjack or roulette. This raises the question if this constitutes a good faith effort not to engage in gambling in New York.[129]

---

[125] 2000 WL 804434 (E.D.Pa. 2000).
[126] See, *e.g*. Expedia <http://www.expedia.com> (date accessed: 31 March 2001).
[127] The legal implications of a mistaken self-declaration is more problematic. The possibility of a mistaken self-declaration is a genuine possibility where the question posed requires a layperson apply legal principles. For example, the answer to "where do you habitually reside?" might differ from the answer to the question "where do you live?". If the consumer is unfamiliar the legal standards for habitual residence, they may mistakenly self-declare their jurisdiction. Under such circumstances, it is unclear whether the consumer should bear the legal burden of the mistaken self-declaration should a dispute arise.

The court's approach to self-declaration is similar to the approach of the court in the iCraveTV case, which, as discussed above, was dismissive of that company's attempts to use contract to limit its signal to Canadians.[130]

Contract must clearly play a central role in any determination of jurisdiction targeting since providing parties with the opportunity to set their own rules enhances legal certainty. As the foregoing review of recent Internet jurisdiction case law reveals, however, contracts do not provide either party with an absolute assurance that their choice will be enforced. Rather, courts must engage in a detailed analysis of how consent was obtained as well as consider the reasonableness of the terms. The results of that analysis should determine what weight to grant the contractual terms when balanced against the remaining two factors of the proposed targeting analysis.

B.      Technology

The second targeting factor focuses on the use of technology to either target or avoid specific jurisdictions. Just as technology originally shaped the Internet, it is now reshaping its boundaries by quickly making geographic identification on the Internet a reality. The rapid emergence of these new technologies challenge what has been treated as a truism in cyberlaw – that the Internet is borderless and thus impervious to attempts to impose on it real-space laws that mirror traditional geographic boundaries.[131]

---

[128] 1999 N.Y. Misc. LEXIS 425 (S.C. N.Y. 1999).
[129] *Id*, ff para. 1.
[130] *iCraveTV*, supra.

Courts have largely accepted the notion that the Internet is borderless as reflected by their reluctance to even consider the possibility that geography might be possible online. In *ALA v. Pataki*, a commerce clause challenge to a New York state law targeting Internet content classified as obscene, the court characterized geography on the Internet in the following manner:

> The Internet is wholly insensitive to geographic distinctions. In almost every case, users of the Internet neither know nor care about the physical location of the Internet resources they access. Internet protocols were designed to ignore rather than document geographic location; while computers on the network do have "addresses," they are logical addresses on the network rather than geographic addresses in real space. The majority of Internet addresses contain no geographic clues and, even where an Internet address provides such a clue, it may be misleading.[132]

Although the ALA court's view of the Internet may have been correct in 1997, the Internet has not remained static. Providers of Internet content increasingly do care about the physical location of Internet resources and the users that access them, as do legislators and courts who may want real space limitations imposed on the online environment.[133] A range of companies have responded to those needs, by developing technologies that provide businesses with the ability to reduce their legal risk by targeting their online presence to particular geographic constituencies and serving the interests of governments and regulators, who may now be better positioned to apply their offline regulations to the online environment.[134]

---

[131] *Law and Borders*, supra.
[132] American Libraries Ass'n v. Pataki (969 F.Supp. 160, 170 (S.D.N.Y.,1997)).
[133] B. Tedeschi, *E-commerce: Borders Returning to the Internet*, New York Times (2 April 2001) online: New York Times < http://www.nytimes.com/2001/04/02/technology/02ECOMMERCE.html> (date accessed: 3 April 2001) [hereinafter *Borders Returning*].
[134] In addition to the discussion below, *see*, Jack L. Goldsmith and Alan O. Sykes, *The Internet and the Dormant Commerce Clause*, 110 Yale L. J. 785, 810-2 (2001).

Since both business and government have a vested interest in bringing geographic borders to the online environment (albeit for different reasons), it should come as little surprise that these technologies have so quickly arrived onto the marketplace. In fact, they have become available before the Internet community has engaged in a discussion on the benefits, challenges, and consequences of creating borders or "zoning" the Internet with these new technologies.[135] This is most unfortunate, since geographic bordering technologies raise important privacy considerations that have, as yet, attracted little debate.[136]

Although critics often point to the inaccuracy of these technologies, few users of the technology actually require perfection.[137] Business wants either to target its message to consumers in a specific jurisdiction or to engage in "jurisdictional avoidance."[138] Effective jurisdictional avoidance provides the means to exclude the majority of visitors who cannot be verified as residing in the desired jurisdiction. For example, iCraveTV did

---

[135] Although in fairness, there are some that saw these developments coming many years ago. For example, Professor Lawrence Lessig, in the same Stanford Law Review issue that featured the Post and Johnson Law and Borders article referred to earlier, commented that:

> In its present design, cyberspace is open, and uncontrolled; regulation is achieved through social forces much like the social forms that regulate real space. It is now unzoned: Borders are not boundaries; they divide one system from another just as Pennsylvania is divided from Ohio. The essence of cyberspace today is the search engine--tools with which one crosses an infinite space, to locate, and go to, the stuff one wants. The space today is open, but only because it is made that way. Or because we made it that way. (For whatever is true about society, at least cyberspace is socially constructed.)
>
> It could be made to be different, and my sense is that it is. The present architecture of cyberspace is changing. If there is one animating idea behind the kinds of reforms pursued both in the social and economic spheres in cyberspace, it is the idea to increase the sophistication of the architecture in cyberspace, to facilitate boundaries rather than borders. It is the movement to bring to zoning to cyberspace.

Lawrence Lessig, *The Zones of Cyberspace*, 48 Stan. L. Rev. 1403, 1408-9 (1996) [hereinafter *Zones*]. See also, *Code*, *supra*, at 56-7.

[136] Stephanie Olsen, *Geographic Tracking Raises Opportunities, Fears*, CNET News.com (8 November 2000); online < http://news.cnet.com/news/0-1005-200-3424168.html> (data accessed: 12 April 2001) [hereinafter *Geographic Tracking*].

[137] As Lessig points out, "A regulation need not be absolutely effective to be sufficiently effective." *Zones* at 1405.
The same applies to bordering technologies; whether used for targeted marketing or to ensure legal compliance, it need not be perfect.

[138] *Borders Returning*, supra.

not use identifying technologies, choosing instead to rely on the user clickwrap agreements.[139] JumpTV, a new Canadian entry into the webcasting market, has indicated that it will use identifying technologies to ensure that only Canadians access its signal.[140] While this may exclude some Canadians who cannot be positively identified as coming from Canada, it will provide the company with a greater level of assurance that it is meeting its goal of limiting its online signal.

Geographic identification technologies can be grouped into at least three categories – a) user identification, which is typically based on IP address identification; b) self-identification, which often occurs through attribute certificates; and c) offline identification.

a.     User Identification

User identification has been utilized on the Internet on a relatively primitive scale for some time. For example, Internet Protocol (IP) lookups, which determine user locations by cross-checking their IP address against databases that list Internet service provider locations, has been used by Microsoft to comply with U.S. regulations prohibiting the export of strong-encryption Web browsers for many years.[141] Although imperfect, the process was viewed as sufficiently effective to meet the standards imposed by the regulations. Recently, several companies have begun offering more sophisticated versions of similar technologies.

i.     Infosplit

---

[139] *iCraveTV*, s*upra*.
[140] Matthew Fraser, *JumpTV Takes On Vested Interests*, Financial Post (29 January 01), online: National Post <http://www.nationalpost.com/search/story.html?f=/stories/20010129/454542.html> (date accessed: 31 March 2001).
[141] A. Jesdanun, *The Potential and Peril of National Internet Boundaries,* S.F. Examiner (March 4, 2001), online: S.F. Examiner <http://www.examiner.com/business/default.jsp?story=b.net.0107> (date accessed: 31 March 2001) [hereinafter *Microsoft*].

Infosplit claims to have the ability to accurately pinpoint the location of any IP address using a proprietary set of techniques and algorithms.[142] The technology provides instant and precise geographic identification and page routing in a process invisible to the Web user. The company maintains that its technology accurately determines the country of origin with 98.5% accuracy, the state or province with 95% accuracy, and the city with 85% accuracy and that it can even accurately determine user location for users of national or global ISPs such as AOL.

The Infosplit technology returns a geographic location by sending the user's IP address to the various algorithms including Trace route, the ARIN/RIPE/APNIC database, and a DNS reverse look-up. The ARIN/RIPE/APNIC component analyzes information obtained from the ARIN/RIPE/APNIC database. The DNS Reverse Lookup component analyzes publicly available domain name registration data. The Trace route algorithm discovers and interprets the trail left by network packets associated with the viewer's Web page request. By combining the results of all three algorithms, Infosplit can provide a more a more effective result that with an IP lookup alone.

ii.     NetGeo

NetGeo is a National Science Foundation project that provides geographic identification primarily through IP address analysis.[143] The project features a database and collection of Perl scripts used to map IP addresses and domain names to geographical locations.

To determine the latitude/longitude values for a domain name, NetGeo first searches for a record containing the target name in its own database. The NetGeo database caches the location information parsed from the results of previous *whois* lookups, to minimize the load on *whois* servers. If a record for the target domain name is found in the database, NetGeo returns the requested information. If no matching record is found in the NetGeo database, NetGeo performs one or more *whois* lookups using the InterNIC and/or RIPE *whois* servers, until a *whois* record for the target domain name is found.

---

[142] See Infosplit <http://www.infosplit.com> (date accessed: 31 March 2001).
[143] See NetGeo <http://www.netgeo.com> (date accessed: 31 March 2001).

After obtaining a record from a *whois* server, the NetGeo Perl scripts parse the *whois* record and extract location information and the date of last update. The NetGeo parser attempts to extract the city, state, and country from the text of the *whois* record. For U.S. addresses the parser also extracts the zip code, if possible. If the parser is unable to parse an address it attempts to find an area code or international phone code in the contact section; the phone code is mapped to a country and then the parser attempts to parse the address again, using the hint provided by the phone code. The parser also guesses the country from email addresses with country-code TLDs found in the contact section.

iii.     EdgeScape

Akamai, a network caching service, also provides a geographic identification service called EdgeScape.[144]  EdgeScape maps user IP addresses to their geographic and network point of origin. This information is assembled into a database and made available to EdgeScape customers.  Each time a user accesses the client's Web site, EdgeScape provides data detailing the country from which user is accessing site, the geographic region within that country (i.e., state or province), and the name of user's origin network.

iv.     Digital Envoy

Founded in 1999, Atlanta-based Digital Envoy's core competency is geographic identification on the Web.[145]  The company's flagship product, NetAcuity, claims country targeting capability exceeding 99% accuracy with targeting of regions, states, or cities also a possibility. The company's primary focus has been the corporate marketing sector, who rely on Digital Envoy to allow for geographically targeted advertising.[146] The company's technology is also used by CinemaNow Inc., a California-based online distributor of feature-length films, which uses the technology to limit distribution of the films to ensure it is compliant with distribution-license rules that vary by country. [147]

---

[144] See Akamai, <http://www.akamai.com/html/en/sv/edgescape_works.html> (date accessed: 31 March 2001).
[145] See Digital Envoy, <http://www.digitalenvoy.com/prod_netacu.htm> (date accessed: 12 April 2001).

v.      Quova

One of the best-funded companies offering geographic identification technologies is Quova,[148] a California-based startup that purchased European leader RealMapping in early 2001.[149] The company spent nine months scanning the Internet's 4.2 billion IP addresses, yielding a detailed physical map of the Internet.[150] The result was the company's flagship product GeoPoint, which boasts 98 percent accuracy at determining Web surfers' countries and 85 percent accuracy on the city level.[151] Currently in development is new technology that will allow for greater identification of AOL users, whose geographic origins are typically more difficult to identify than most other ISPs.[152]

b.      Self-identification

Unlike user identification technologies, which identify the user's geographic location without requesting permission from the user to do so, self-identification uses technologies that enable users to provide geographic identification directly to the Web site. This is most frequently accomplished through the use of attribute certificates, which, as Professor Michael Froomkin explains, provide information about the attributes of a particular user without revealing their actual identity.

---

[146] Nicole Harris, *Digital Envoy Offers a Way To 'Geo-Target' Web Surfers*, Wall Street Journal Interactive (12 April 2001); online <http://interactive.wsj.com/articles/SB987024862803391097.htm> (date accessed: 12 April 2001).
[147] Patricia Jacobus, *CinemaNow Appeases Studios By Locating Web Surfers,* CNET News.com (26 February 2001); online < http://news.cnet.com/news/0-1005-200-3424168.html> (date accessed: 12 April 2001).
[148] See Quova, <http://www.quova.com> (date accessed: 12 April 2001).
[149] Stephanie Olsen, *Tracking Web Users into European Territory,* CNET News.com (3 April 2001); online <http://news.cnet.com/news/0-1005-200-5461873.html> (date accessed: 12 April 2001) [hereinafter *Web Users*].
[150] *Geographic Tracking*, *supra*.
[151] *Web Users, supra.*
[152] *Web Users, supra*.

> Although identifying certificates are likely to be the most popular type of certificate in the short run, in the medium term CAs are likely to begin certifying attributes other than identity. An authorizing certificate might state where the subject resides, the subject's age, that the subject is a member in good standing of an organization, that the subject is a registered user of a product, or that the subject possesses a license such as bar membership. These authorizing certificates have many potential applications. For example, law professors exchanging exam questions on the Internet could require that correspondents demonstrate their membership in the Association of American Law Schools (AALS) before being allowed to have a copy of the questions.
>
> It is illegal to export high-grade cryptography from the United States without advance permission from the federal government, but there are no legal restrictions on the distribution of strong cryptography to resident aliens or United States citizens in the United States. The lack of a reliable means to identify the geographical location of a person from an Internet address creates a risk of prosecution for anyone making cryptographic software available over the Internet. For example, if Alice is making high-grade cryptography available for distribution over the Internet, she might protect herself from considerable risk by requiring that Bob produce a valid certificate from a reputable CA, stating that he is a United States citizen or green card holder residing in the United States, before allowing him to download the cryptographic software.
>
> Alice substantially reduces her risk under the ITAR by requiring Bob to produce an authorizing certificate demonstrating his citizenship, but even this does not eliminate her risk. Alice's major remaining risks are that: (1) the CA's statement was erroneous; (2) Bob has lost control of his digital signature and it has fallen into the hands of Mallet, who is not a United States citizen or permanent resident, or is abroad; and (3) something about Bob has changed since he procured the certificate, for example, he has moved abroad, lost his citizenship or green card, or has died and his private key is held by his executor or heir.
>
> A certificate binding the geographic location, age, or other attribute to a public key can contain the name of the subject of the certificate, but the public key suffices if it was generated in a secure manner and is sufficiently long to be unique. Nameless, anonymous certificates create the possibility for sophisticated anonymous Internet commerce. For example, persons wishing to purchase materials that can only be sold to adults might obtain over 18 certificates that bind this attribute to a public key but do not mention their name. Similarly, a financial institution might issue a certificate linking a public key to a numbered deposit account.[153]

---

[153] A. Michael Froomkin, *The Essential Role of Trusted Third Parties in Electronic Commerce,* 75 Oregon L. Rev. 49 (1996).

Self-identification technology represents a middle ground between user identification, which puts the power of identification solely in the hands of the Web site, and self-declaration, in which the user declares where they reside but without any independent or technological verification of the accuracy of the declaration. The danger with self-identification technologies is that if they become popular, they may also quickly cease to be voluntary since businesses may begin to require that their users supply the data contained in an attribute certificate in order to obtain service.[154]

c.     Offline Identification

Offline identification combines an online presence with certain offline knowledge to form a geographic profile of a user. The best example of offline identification is credit card data. Since credit cards remain the preferred payment mechanism for most online transactions, sellers are regularly asked to verify the validity of a user's credit card. As anyone who has purchased online with a credit card knows, the verification process includes an offline component, as the address submitted by the user is cross-checked with the address on file to confirm a match prior to authorization of the charge.[155] This process provides Web sites with access to offline data such as the user's complete address – which is confirmed through a third party, the financial intermediary.

While this system may be effective for sites that are actively engaged in e-commerce and for those whose geographic risks are confined strictly to those circumstances when they are selling into a particular jurisdiction, the use of credit-card data is of limited utility to those who do not actively sell online or those who are concerned about jurisdictional issues prior to the submission of a credit card number and address information.

---

[154] *Code*, *supra*, at 42.
[155] *Credit Card Fraud Crippling Online Merchants*, E-commerce Times (20 March 2000), online: E-commerce Times <http://www.ecommercetimes.com/news/articles2000/000320-2.shtml> (date accessed: 31 March 2001) ("At present, credit card companies only verify if a credit card number is correct and then match the number against the customer's billing address.").

Two other offline identifiers present similar possibilities of geographic identification, but simultaneously raise serious privacy concerns. At one time, Microsoft included a feature in its software that could be used to transmit personal information via the Internet without the user's knowledge. The feature enabled Microsoft software such as Word or Excel to issue identification numbers unique to the software and the computer on which the software was installed. During the online registration process, the number, known as a Global Unique Identifier (GUID), was transferred to Microsoft along with the user's name, address, and other personal information. Microsoft could then identify users by matching their GUID with the information stored in the Microsoft-controlled database.[156]

Intel found itself embroiled in a similar privacy controversy when it was revealed that the company was able to identify people online by using a Processor Serial Number (PSN), which was a number burned onto an Intel processor chip at the time of manufacture and designed to provide authentication in Internet communication and commerce. The PSN identified a person on the Internet by the actual hardware of the computer they were using, limiting their ability to be anonymous on the Web, and providing for the prospect of quick identification.[157]

Though clearly limited in scope, offline identifiers may often provide the most inexpensive method of identifying geographic location since they rely on offline data that is collected independently of online activities. Precisely because they merge offline and online, these technologies raise profound privacy concerns, creating the prospect of personally identifiable information being transferred along with non-identifiable geographic data.

d.     Targeting and Technology

---

[156] M. Ricciuti, *Microsoft Admits Privacy Problem, Plans Fix,* CNET News.com (7 March 1999), online: CNET News.com <http://news.cnet.com/news/0-1006-200-339622.html> (date accessed: 31 March 2001).

[157] *Privacy and the Internet Tutorial: Protecting Web Privacy - Processor Serial Number*, Privacy Exchange, online: Privacy Exchange <http://www.privacyexchange.org/tsi/psn.htm> (date accessed: 31 March 2001).

Given the development of new technologies that allow for geographic identification with a reasonable degree of accuracy, a targeting test must include a technology component that places the onus on the party contesting or asserting jurisdiction to demonstrate what technical measures it employed to either target or avoid a particular jurisdiction.  The suitability of such an onus lies in the core consideration of jurisdiction law – that is, whether jurisdiction is foreseeable under the circumstances.  Geographic identifying technologies provide the party that deploys the technology with a credible answer to that question.  Since parties can identify who is accessing their site, they can use technical measures to stop people from legally-risky jurisdictions from doing so. A fair and balanced targeting jurisdiction test demands that they do just that.

It is important to note that parties are not typically required to use geographic identification technologies.[158]  In many instances, they do not care who accesses their site and thus will not be willing to incur the expense of installing such systems.  In other instances, the party may be acutely aware of the need to identify users from a jurisdiction that bans access to certain content or certain activities.  In such instances, the party may wish to limit access to those users it can positively identify from a legally-safe jurisdiction.

The inclusion of technology into the targeting test does not, therefore, obligate parties to use the technology.  Rather, it forces parties to acknowledge that such technologies are available and that prudence may dictate using them in some capacity.  Moreover, the test does not prescribe any specific technology – it only requires that consideration be given to the technologies used and available at a particular moment in time.   This technology neutral prong of the targeting test also provides an effective counter-balance to the contract and knowledge factors.  It removes the ability to be willfully blind to users who enter into a clickwrap contract stating that they are from one jurisdiction, while the technological evidence suggests something else entirely.

---

[158] Except where as required by law.  *See*, *e.g*. *Microsoft*, *supra*.

C. Actual or Implied Knowledge

The third targeting factor assesses the knowledge the parties had or ought to have had about the geographic location of the online activity. Although some authors have suggested that the Internet renders intent and knowledge obsolete by virtue of the Internet's architecture,[159] the geographic identification technologies described above do not support this view.

In certain respects, this factor is little more than an extension of the contract and technology factor. It seeks to ensure that parties cannot hide behind contracts and/or technology by claiming a lack of targeting knowledge when the evidence suggests otherwise.

The implied knowledge factor is most apparent in the defamation tort cases that follow from the *Calder* decision. In those cases, courts have accepted that the defaming party is or should be aware that the injury inflicted by their speech would be felt in the jurisdiction of their target. Accordingly, in such cases a party would be unable to rely on a contract that specifies an alternate jurisdiction as the choice of forum.

The court's desire to dismiss any hint of wilfull blindness is evident in the *People v. World Interactive Gaming* case, referred to earlier.[160] In that case, the online casino argued that it had limited access to only those users that had entered an address of a jurisdiction where gambling was permitted. The court saw through this ruse, however, firmly stating that:

---

[159] See, *e.g.,* Martin H. Redish, *Of New Wine and Old Bottles: Personal Jurisdiction, The Internet, and the Nature of Constitutional Evolution*, 38 Jurimetrics J. 575, 605-6(1998). ("the most effective defense of an Internet exception to the purposeful availment requirement is not that state interest should play an important role only in Internet cases, but rather that the technological development of the Internet effectively renders the concept of purposeful availment both conceptually incoherent and practically irrelevant. An individual or entity may so easily and quickly reach the entire world with its messages that it is simply not helpful to inquire whether, in taking such action, that individual or entity has consciously and carefully made the decision either to affiliate with the forum state or seek to acquire its benefits.")
[160] *World Interactive Gaming*, *supra.*

> [t]his Court rejects respondents' argument that it unknowingly accepted bets from New York residents. New York users can easily circumvent the casino software in order to play by the simple expedient of entering an out-of-state address. Respondents' violation of the Penal Law is that they persisted in continuous illegal conduct directed toward the creation, establishment, and advancement of unauthorized gambling. The violation had occurred long before a New York resident ever staked a bet. Because all of respondents' activities illegally advanced gambling, this Court finds that they have knowingly violated Penal Law @ 225.05.[161]

The relevance of a knowledge-based factor extends beyond reliance on contracts that the parties know to be false. In an e-commerce context, the knowledge that comes from order fulfillment is just as important. For example, sales of physical goods such as computer equipment or books, provides online sellers with data such as a real-space delivery address, making it relatively easy to exclude jurisdictions that the seller does not wish to target.

Although the application of this principle is more complex when the sale involves digital goods for which there is no offline delivery, the seller is still customarily furnished with potentially relevant information. As discussed above, most telling may be credit card data that the purchaser typically provides to the seller. In addition to the credit card number and expiry data, the purchaser is often also required to supply billing address information so that the validity of the card can be verified before authorization. Since the seller is supplied with a real-space billing address for digital transactions, there remains the opportunity to forego the sale if there is a jurisdictional concern. For example, the Washington Capitals hockey team recently rejected attempts by rival fans from Pittsburgh to purchase tickets on the team's Web site. The site was set to reject purchase attempts from customers entering a Pittsburgh-area phone area code.[162] While some sellers may be loath to use consumer payment information in this fashion, the approach reflects a more general trend toward recognizing the important role that payment

---

[161] *Id.*

[162] Thomas Heath, *Capitals Owner Puts Pittsburgh Fans on Ice*, Washington Post (14 April 2001); online <http://www.washingtonpost.com/wp-dyn/sports/leaguesandsports/nhl/19992000/washingtoncapitals/A16569-2001Apr13.html> (date accessed: 15 April 2001).

intermediaries such as credit card companies play in the consumer e-commerce process.[163]

**Part V – Conclusions**

With courts increasingly resisting the *Zippo* passive versus active approach to Internet jurisdiction, the time for the adoption of a new targeting-based test has arrived. Unlike the *Zippo* test, which suffers from a series of drawbacks including inconsistent and undesirable outcomes as well as the limitations of a technology-specific approach, a targeting-based analysis provides all interested parties – including courts, e-commerce companies, and consumers -- with the tools needed to conduct more effective legal risk analysis.

Under the three-factor targeting test, it is important to note that no single factor is determinative. Analysis will depend on a combined assessment of all three factors in order to determine whether the party knowingly targeted the particular jurisdiction and could reasonably foresee being hauled into court there. In an e-commerce context, the targeting test ultimately establishes a trade-off that should benefit both companies and consumers. Companies benefit from the assurance that operating an e-commerce site will not necessarily result in jurisdictional claims from any jurisdiction worldwide. They can more confidently limit their legal risk exposure by targeting only those countries where they are compliant with local law. Consumers also benefit from this approach since they receive the reassurance that online companies that target them will be answerable to their local law.

---

[163] In March 2001, the *Electronic Commerce and Information, Consumer Protection Amendment and Manitoba Evidence Amendment Act (S.M. 200, c. E55. 77)* and the *Internet Agreements Regulations (Man. Reg. 176/2000)* took effect within the province. Designed to foster an online environment where consumer confidence will flourish, the new laws apply exclusively to the online retail sale of goods or services or the retail lease-to-own of goods between buyers and sellers. Under the new rules, binding e-commerce transactions require the seller to provide certain obligatory information to the buyer under threat of a purchaser contract cancellation remedy.

The test is sufficiently flexible to allow companies to deploy as many or as few precautions as needed. For example, if the company is involved in a highly regulated or controversial field, it will likely want to confine its activities to a limited number of jurisdictions, avoiding locations with which it is not familiar. Under the targeting test, the company could adopt a strategy of implementing technological measures to identify its geographic reach, while simultaneously incorporating the desired limitations into its contract package. Conversely, companies with fewer legal concerns and a desire to sell worldwide without regard for borders, can still accomplish this goal under the targeting test analysis. These companies would sell without the technological support, incurring both the benefits and responsibilities of a global e-commerce enterprise.

Notwithstanding the advantages of a targeting test, there are, nevertheless, some potential drawbacks. First, the test accelerates the creation of a bordered Internet. Although a bordered Internet carries certain advantages, it is also subject to abuse, since countries can use bordering technologies to keep foreign influences out and suppress free speech locally. Second, the targeting test might also result in less consumer choice since many sellers may stop selling to consumers in certain jurisdictions where its risk analysis suggests that the benefits are not worth the potential legal risks.

The most effective illustration of the advantages of a targeting test comes from considering how the test would apply to the two cases outlined at the start of this paper. Although the outcomes would remain the unchanged, the analysis would be different, providing courts and companies alike with a clearer sense of where the boundaries lie.

In the Yahoo! France case, the company argued vociferously that it should not be subject to the jurisdiction of the French court because its flagship dot-com site had not targeted France. Applying the targeting test's three factors, however, suggests that the French court handled the case correctly. Yahoo! utilizes a terms and conditions page stipulating that the site is governed by U.S. law, but as the *Ticketmaster* case demonstrated, that form of contract may not be enforceable. Moreover, the company did not employ any form of technological measures to identify the geographic location of visitors accessing its site and it was likely aware that some visitors were French residents. While the outcome may be the same, the use of a targeting test would have provided the company with a more effective tool to gauge the likelihood of a foreign court asserting jurisdiction.

Applying the targeting test to the iCraveTV case, the U.S. court might still have asserted jurisdiction over the company, but it would have done so for different reasons. Using a targeting analysis, the company would have pointed to its self-declaration contracts in which users affirmed that they were resident in Canada as well as its (rather porous) technological measures that were designed to keep its signal within Canada.  A court would have likely reviewed the iCraveTV effort and, noting that the company was well aware that U.S. residents were accessing the site and that its technical measures were ineffective, would have asserted jurisdiction.  The case also highlights how a successor to iCraveTV could effectively limit its jurisdictional liability by employing stronger technological measures to keep its signal from straying across the border.

Although the targeting test will not alter every jurisdictional outcome, it will provide all parties with greater legal certainty and a more effective means of conducting legal risk assessments.  The move toward using contract and technology to erect virtual borders may not answer the question of whether there is a there there, but at least it will go a long way in determining where the there might be.